\documentclass[11pt,onecolumn]{IEEEtran}

\usepackage{enumerate}
\usepackage[utf8]{inputenc} 
\usepackage[T1]{fontenc}
\usepackage{url}
\usepackage{ifthen}
\usepackage{tcolorbox}
\usepackage[noadjust]{cite}
\usepackage[cmex10]{amsmath} \usepackage{amssymb,amsfonts,amscd,amsbsy}
\usepackage{graphicx}
\usepackage{stmaryrd} 
\usepackage[mathscr]{eucal}
\usepackage{mathtools}
\usepackage{bbm}
\usepackage{xcolor}
\usepackage{algorithm}
\usepackage{algpseudocode}
\usepackage{enumerate}

\usepackage{hyperref}
\definecolor{cardinal}{rgb}{0.77, 0.12, 0.23}
\hypersetup{
    colorlinks=true,
    linkcolor=cardinal,
    citecolor=cardinal,
    urlcolor=cardinal
    }

\newtheorem{corollary}{Corollary}
\newtheorem{lemma}{Lemma}

\newtheorem{claim}{Claim}
\newtheorem{example}{Example}
\newtheorem{remark}{Remark}

\usepackage{etoolbox}
\AtEndEnvironment{example}{\null\hfill$\square$}%
\AtEndEnvironment{remark}{\null\hfill$\square$}%

\newcommand{\suppress}[1]{}

\newcommand{\Fb}{\mathbbmss{F}} 
\newcommand{\lset}{\llbracket}  
\newcommand{\rset}{\rrbracket}  
\newcommand{\Cs}{\mathscr{C}}   
\newcommand{\dber}[3]{\mathsf{DB}_{#1}(#2, #3)} 

\newcommand{\Amin}{\mathcal{A}_{\min}}   
\newcommand{\Ps}{\mathscr{P}}   
\newcommand{\wt}{w_H}   
\newcommand{\supp}{\mathsf{supp}}  

\newcommand{\RM}{\mathrm{RM}}   
\newcommand{\tth}{\text{th}}    
\newcommand{\p}{\pmb}           

\newcommand{\Hs}{\mathscr{H}}

\newcommand{\dmin}{d_{\min}}    

\newcommand{\nosubscriptallone}{\mathbbm{1}}   
\newcommand{\allonelengthn}{\mathbbm{1}_n}   
\newcommand{\dimenof}{\mathsf{dim}}  
\newcommand{\spanof}{\mathsf{span}}  
\newcommand{\rowspaceof}{\mathsf{rowsp}}  

\newcommand\code[2]{\mathscr{C}^{\otimes[#1,#2]}} 

\newcommand{\order}{\mathscr{O}}

\newcommand{\Csub}{\mathscr{C}_{sub}} 

\newcommand{\alloneB}{\mathbbm{1}_{n^{(m-1)}}}
\newcommand{\Rb}{\mathbbmss{R}}
\newcommand{\Pb}{\mathbbmss{P}}

\newcommand{\Fs}{\mathscr{F}}
\newcommand{\Vn}{{\sf V}}
\newcommand{\Vnh}{{\sf V}_{\sf h}}
\newcommand{\Chn}{{\sf C}}
\newcommand{\Chng}{{\sf C}_{\sf g}}
\newcommand{\nb}{\partial}
\newcommand{\gammag}{\gamma_{\sf g}}
\newcommand{\step}{\textsc{Step}}

\title{Recursive Subproduct Codes\\ with Reed-Muller-like Structure}

\author{Aditya Siddheshwar, Lakshmi Prasad Natarajan, Prasad Krishnan%
\thanks{\hrule}%
\thanks{Aditya Siddheshwar and Prasad Krishnan are with the Signal Processing and Communications Research Center, International Institute of Information Technology, Hyderabad 500032, India (email: aditya.siddheshwar@research.iiit.ac.in, prasad.krishnan@iiit.ac.in).}%
\thanks{Lakshmi Prasad Natarajan is with the Department of Electrical Engineering, Indian Institute of Technology Hyderabad, Sangareddy 502284, India (email: lakshminatarajan@iith.ac.in).}
\thanks{This work was supported by the Qualcomm 6G University Research India Program.}
}

\begin{document}

\maketitle 

\begin{abstract}
We study a family of subcodes of the $m$-dimensional product code $\mathscr{C}^{\otimes m}$ (`subproduct codes') that have a recursive Plotkin-like structure, and which include Reed-Muller (RM) codes and Dual Berman codes as special cases. We denote the codes in this family as $\mathscr{C}^{\otimes [r,m]}$, where \mbox{$0 \leq r \leq m$} is the `order' of the code. These codes allow a `projection' operation that can be exploited in iterative decoding, viz., the sum of two carefully chosen subvectors of any codeword in $\mathscr{C}^{\otimes [r,m]}$ belongs to $\mathscr{C}^{\otimes [r-1,m-1]}$. Recursive subproduct codes provide a wide range of rates and block lengths compared to RM codes while possessing several of their structural properties, such as the Plotkin-like design, the projection property, and fast ML decoding of first-order codes. Our simulation results for first-order and second-order codes, that are based on a belief propagation decoder and a local graph search algorithm, show instances of subproduct codes that perform either better than or within $0.5$~dB of comparable RM codes and CRC-aided Polar codes.
\end{abstract}

\setcounter{tocdepth}{2} 
\tableofcontents
\clearpage


\section{Introduction}




The product construction~\cite{Elias_IRE_54} is a well-known technique to obtain new codes from existing codes.
When used $m$ times repeatedly on an $[n,k,d]$ code $\Cs$ we obtain an $m$-fold product code $\Cs^{\otimes m}$ with parameters $[n^m,k^m,d^m]$. 
The parity-check constraints of product codes can be expressed through a factor graph with {generalized check nodes}. 
Hence, it is natural to use iterative techniques to decode product codes~\cite{Tan_IT_81,Justesen_11}.
The product construction has been a popular technique for designing coding schemes for the additive white Gaussian noise (AWGN) channel; examples of some recent work in this direction include~\cite{CJL_COMML_19,JFMH_ICC_22,CLGLP_IT_23}.

Several variants of the product construction are known that have yielded codes with good error correction properties. 
Codes intended for use in optical transport networks have been designed from subcodes of product codes, or \emph{subproduct codes}, that satisfy certain symmetry properties~\cite{Justesen_11,MPPP_ISIT_15,PKN_ITA_15}. Ideas from convolutional coding have been combined with product construction to design braided block codes~\cite{FTLZ_IT_09} and staircase codes~\cite{SFHKL_Lightwave_12}.

The objective of this work is to use the product construction to identify codes that satisfy a \emph{projection} property: the sum of two carefully chosen subvectors of any codeword lies in another code for which a fast decoder is available.
In such cases, the latter code can be used as a generalized check node in the factor graph for iterative decoding. 
Reed-Muller (RM) codes~\cite{Muller_IRE_54,Reed_IRE_54} enjoy such a projection property, viz., a projection applied on $\RM(r,m)$ yields $\RM(r-1,m-1)$. 
This was used to design the recursive projection aggregation (RPA) decoder for RM codes in~\cite{YeA_IT_20} and a related belief propagation (BP) decoder in~\cite{LHP_ISIT_20} by using the fact that there exists a fast maximum-likelihood (ML) decoder for $\RM(1,m)$ codes~\cite{BeS_IT_86,AsL_IT_04,ASY_IT_20}.
A similar decoding technique for extended cyclic codes was proposed in~\cite{ZhH_Globecom_22}.

Starting from any \mbox{$[n,k \geq 2,d]$} code $\Cs$ which contains the all-one codeword, we identify a family of subproduct codes 
\begin{equation*}
\{\nosubscriptallone,\p{0}\} = \code{0}{m} \subset \code{1}{m} \subset \cdots \subset \code{m}{m} = \Cs^{\otimes m}.
\end{equation*}
The parameters of $\code{r}{m}$, $r \in \{0,1,\dots,m\}$, are 
$$\left[ n^m, \sum_{l=0}^{r}\binom{m}{l}(k-1)^l,d^{\,r} n^{(m-r)}\right].$$
These codes share several structural properties with RM codes, and in fact, include the RM codes and Dual Berman codes~\cite{BlN_IT_01,Ber_Cybernetics_II_67,NaK_IT_23} as special cases, corresponding to, $\Cs=\Fb_2^2$ and $\Cs=\Fb_2^n$, $n \geq 3$, respectively. 
These subproduct codes satisfy a recursive structure which generalizes the Plotkin construction of RM codes.
As with RM codes, a projection operation can be applied on $\code{r}{m}$ to yield $\code{r-1}{m-1}$, and further, the code $\code{1}{m}$ can be ML decoded efficiently.  

We describe the construction and basic properties of the recursive subproduct codes  $\code{r}{m}$, including the identification of their minimum weight codewords, in~Section~\ref{sec:recursive_subproduct_codes}.
A fast ML decoder and a soft-output max-log-MAP decoder for the first-order codes $\code{1}{m}$ is described in Section~\ref{sec:first_order_decoding}. 
In Section~\ref{sec:second_order_decoding} we introduce the projection operation, describe a BP decoder that uses projections, and an improvement of this BP decoder using a local graph search algorithm~\cite{Kam_TCOM_22} for second-order codes $\code{2}{m}$. 
%
We present some simulation results for first- and second-order codes in Section~\ref{sec:simulations} to show that it is possible to design subproduct codes that perform either better than or within $0.5$~dB of comparable RM codes and CRC-aided Polar codes.

Recursive subproduct codes provide a wider range of rates and block lengths compared to RM codes while possessing several of their structural properties.
We believe that these codes warrant further investigation, especially regarding their applicability in low-capacity channels~\cite{FHJM_JSAIT_23},
designing more efficient decoders, 
and improving the performance using high-rate outer codes. 
It is likely that these codes have applications in other areas where RM codes or Dual Berman codes are useful, such as in private information retrieval~\cite{PIR_Transitive_Codes,PIR_Berman}.
We conclude this paper in Section~\ref{sec:discussion} where we have included remarks about a few natural approaches that did not seem to work in our simulation experiments.

\emph{Notation:} The symbol $\otimes$ denotes the Kronecker product. For any positive integer $\ell$, let $\lset \ell \rset$ denote the set $\{0,1,\dots,\ell-1\}$.
The binary field is $\Fb_2=\{0,1\}$. 
The empty set is $\emptyset$.
For sets $A, B$ we define $A \setminus B = \left\{a \in A~|~ a \notin B \right\}$. 
We use capital letters to denote matrices (such as $G$), and small bold letters to denote row vectors (such as $\p{g}$).
The notation $(.)^T$ denotes the transpose operator.
The notation $\p 0$ denotes a zero-vector of appropriate size. 
For two vectors ${\p a},{\p b}$, their concatenation is denoted by $({\p a}|{\p b}).$  The dimension of a code $\mathcal{C}$ is denoted by $\dimenof(\mathcal{C})$ and its minimum distance by $\dmin(\mathcal{C})$. The Hamming weight of a vector ${\p a}$ is denoted by $\wt({\p a})$. 
We use $\spanof$ to denote the span of a collection of vectors, and $\rowspaceof$ to denote the span of the rows of a matrix.
The binomial coefficient $\binom{n}{k}$ is assumed to be $0$ if $k>n$ or if $k<0$.  

\section{Recursive Subproduct Codes} \label{sec:recursive_subproduct_codes}

Denote the all-one row vector in $\Fb_2^n$ by $\allonelengthn$. Let $k,n$ be positive integers such that \mbox{$2\leq k\leq n$}. Consider an $[n,k,d]$ code $\Cs$, such that \mbox{$\allonelengthn\in\Cs$}. 
Towards providing the construction of the recursive subproduct codes, 
we choose a largest subcode $\Csub$ of $\Cs$ such that \mbox{$\allonelengthn\notin\Cs_{sub}$}. 
We will show later (in Claim \ref{claim:csubdoesntmatter}) that the specific choice of such a code $\Cs_{sub}$ does not affect our construction.
It is easy to see that \mbox{$\dimenof(\Cs_{sub})=k-1$} and \mbox{$\Cs=\Cs_{sub}\bigoplus \spanof(\{\allonelengthn\})$}.
Let $G_{sub}=\begin{bmatrix}
\p{g}_1^T&
\p{g}_2^T&
\hdots&
\p{g}_{k-1}^T
\end{bmatrix}^T$ denote a generator matrix of $\Cs_{sub}$.  Then the generator matrix of $\Cs$ can be obtained as $\begin{bmatrix}{\p g}_0^T&G_{sub}^T\end{bmatrix}$ where we use ${\p g}_0=\allonelengthn$. Let $m>0$ be an integer. For a tuple $\p j=(j_0,\hdots,j_{m-1})\in \lset k \rset^m$, define the vector 
${\p b}_{\p j} \in \Fb_2^{\,n^m}$ 
as follows,
\begin{align}
    {\p b}_{\p j}
    =\p{g}_{j_0}\otimes \p{g}_{j_1}\otimes \cdots\otimes \p{g}_{j_{m-1}}. \label{eq:bj_basis_vector}
\end{align}

For any $0\leq l \leq m$, let $B_{l,m}$ denote the matrix of size $\binom{m}{l}(k-1)^l \times n^m$, with rows indexed by $\{{\p j}\in \lset k \rset^m: \wt({\p j})=l\}$ in lexicographic order, such that the ${\p j}^{\tth}$ row of $B_{l,m}$ is precisely 
${\p b}_{\p j}$. 
Note that $B_{0,m}=\begin{bmatrix}b_{(0,0,0,0,\cdots,0)}\end{bmatrix}=\begin{bmatrix}\nosubscriptallone_{n^m}\end{bmatrix}$. As another example, we have $B_{2,m}$ as the following matrix
\begin{align*}
\begin{bmatrix}
\p b_{(1,1,0,0,\cdots,0)}\\
\p b_{(1,2,0,0,\cdots,0)}\\
\vdots\\
\p b_{(0,0,\cdots,0,\cdots,k-1,k-2)}\\
\p b_{(0,0,\cdots,0,\cdots,k-1,k-1)}
\end{bmatrix}
&=\begin{bmatrix}
{\p g}_1\otimes{\p g}_1\otimes\allonelengthn\otimes\cdots\otimes \allonelengthn \\
{\p g}_1\otimes{\p g}_2\otimes\allonelengthn\otimes\cdots\otimes \allonelengthn \\
\vdots\\
\allonelengthn\otimes\cdots\otimes\allonelengthn\otimes{\p g}_{k-1}\otimes {\p g}_{k-2} \\
\allonelengthn\otimes\cdots\otimes\allonelengthn\otimes{\p g}_{k-1}\otimes {\p g}_{k-1}
\end{bmatrix}.
\end{align*}

For $0\leq r\leq m$, we now define the matrices $G_{r,m}$ as 
\begin{align}
\label{eqn:genmatrixfortensorproductcode}
    G_{r,m}=
    \begin{bmatrix}
    B_{0,m}\\
    B_{1,m}\\
    \vdots\\
    B_{r-1,m}\\
    B_{r,m}
    \end{bmatrix}. 
\end{align}
We denote by $\code{r}{m}$ the $n^m$-length linear code obtained as the rowspace of the matrix $G_{r,m}$. That is,
\begin{equation*}
\code{r}{m} = \spanof\left( \left\{ 
\p{b}_{\p{j}}: \p{j} \in \lset k \rset^m, \wt(\p{j}) \leq r
\right\} \right).
\end{equation*}
We define $\code{r}{m} = \{\p 0\}$ if $r<0$ or $r>m$. 
We will refer to $r$ as the \textit{order} of the code $\code{r}{m}$.
\begin{example}
\label{example:order0andorderm}
By definition, $G_{0,m}$ is $\begin{bmatrix}\nosubscriptallone_{n^m}\end{bmatrix}$. Hence $\code{0}{m}$ is just the repetition code of length $n^m$. Similarly, we observe from the structure of the matrix $G_{m,m}$ that $G_{m,m}=G^{\otimes m}$. Thus, $\code{m}{m}=\Cs^{\otimes m}$ is the $m$-fold product code of the code $\Cs$. 
Note that $\code{r}{m} \subseteq \code{m}{m} = \Cs^{\otimes m}$ for every $r \leq m$.
\end{example}

\begin{remark}
We note that~\eqref{eqn:genmatrixfortensorproductcode}
can be expressed recursively as 
\begin{equation*}
G_{r,m} = 
P \begin{bmatrix}
\allonelengthn \otimes G_{r,m-1} \\
G_{sub} \otimes G_{r-1,m-1}
\end{bmatrix},
\end{equation*}
for some permutation matrix $P$.
Lemma~\ref{lemma:decompositionofCsrm} describes the resulting recursive structure of $\code{r}{m}$.
Although we do not study the dual code of $\code{r}{m}$ in this paper, we remark that the parity-check matrix of the dual code, viz., $P^{-1}G_{r,m}$, is a block matrix composed of Kronecker products. Codes with such matrices as their parity check matrices have been explored in \cite{Wolf_ITW_06,Imai_Fujiya_IT_81,Binary_LRC}, under the name of \textit{generalized tensor product codes}. The dual code of $\code{r}{m}$ is therefore such a generalized tensor product code.
\end{remark}

\subsection{Basic Properties of Recursive Subproduct Codes}

\begin{lemma} \label{lemma:dimension_of_new_code}
    $\dimenof(\code{r}{m})=\sum_{l=0}^r\binom{m}{l}(k-1)^l$. 
\end{lemma}
\begin{IEEEproof}
    Note that $G_{r,m}$ is a submatrix of $G_{m,m}=G^{\otimes m}$. Now, as $G$ has rank $k$, and thus the rank of $G^{\otimes m}$ is $k^m=(1+(k-1))^m=\sum_{l=0}^m\binom{m}{l}(k-1)^l$, which is the number of rows of $G^{\otimes m}$. Thus, $G_{r,m}$ has full row-rank as well, which completes the proof. 
\end{IEEEproof}
\begin{lemma}
\label{lemma:decompositionofCsrm}
\emph{(Recursive Plotkin-like structure of $\code{r}{m}$)}
\begin{align}
\label{eqn:codewordofCsrmdecomposition}
\code{r}{m}&=\rowspaceof(G_{r,m}) 
=\Big\{\sum_{i=0}^{k-1}{\p d}_i\otimes {\p g}_i\colon \forall {\p d}_0\in\code{r}{m-1}, 
\forall {\p d}_i\in\code{r-1}{m-1}, \forall i\in \{1,\hdots,k-1\} \Big\}.
\end{align}
Further, for each codeword in $\code{r}{m}$, there exists a unique choice for the codewords ${\p d}_i: i \in \lset k \rset$ in the decomposition in R.H.S. of (\ref{eqn:codewordofCsrmdecomposition}). 
\end{lemma}
\begin{IEEEproof}
Please see Appendix~\ref{app:proof:lemma:decompositionofCsrm}.
\end{IEEEproof}

The following claim shows that the code $\code{r}{m}$ is invariant to the choice of the $(k-1)$-dimensional code $\Cs_{sub}$, as long as it does not contain the codeword $\allonelengthn$. 
\begin{claim}
\label{claim:csubdoesntmatter}
Let $\Csub'$ denote any $(k-1)$-dimensional subcode of $\Cs$ which does not contain the codeword $\allonelengthn$. Let $G_{sub}'$ denote a generator matrix of $\Csub'$. Consider the matrix $G'_{r,m}$ defined as in the R.H.S. of (\ref{eqn:genmatrixfortensorproductcode}), using the code  $G_{sub}'$ (instead of $G_{sub}$). Then, $\rowspaceof(G'_{r,m})=\code{r}{m}$. 
\end{claim}
\begin{IEEEproof}
We prove the claim by induction on $m$. 
For $r=m$ (respectively $r=0$), the matrix $G_{r,m}$ as defined by (\ref{eqn:genmatrixfortensorproductcode}) generates $\Cs^{\otimes m}$ (respectively, the repetition code of length $n^m$), for any subcode $\Csub$ not containing $\allonelengthn$ (see Example \ref{example:order0andorderm}). This also means that the claim is true for $m=1$, for each $r\in \{0,1\}$.

Now assume that the claim is true for $m-1$ and for each value of $r \in \{0,1,\dots,m-1\}$.
We now prove that it is true for $m$ as well. Clearly, it is sufficient to consider $r\in \{1,\hdots,m-1\}$. 

Let 
$G_{sub}'=\begin{bmatrix}
    \p{\tilde{g}}_1^T&
    \p{\tilde{g}}_2^T&
    \hdots&
    \p{\tilde{g}}_{k-1}^T
    \end{bmatrix}^T$ be the given generator matrix for $\Cs_{sub}'$ and $\p{\tilde{g}}_0 = \allonelengthn$. Thus, $G'=\begin{bmatrix}
    \allonelengthn \\
    G_{sub}'
    \end{bmatrix}$ is another generator matrix for $\Cs$. 
    
    Thus, each row of the generator matrix $G=\begin{bmatrix}
    \allonelengthn^T&
    \p{g}_1^T&
    \p{{g}}_2^T&
    \hdots&
    \p{{g}}_{k-1}^T
    \end{bmatrix}^T$ of $\Cs$ can be obtained as a linear combination of the rows in $G'$. Recall that $G$ generates $\Cs$, and $G_{sub}=\begin{bmatrix}
    \p{g}_1^T&
    \p{{g}}_2^T&
    \hdots&
    \p{{g}}_{k-1}^T
    \end{bmatrix}^T$ generates $\Csub$. 
    Therefore, for any $i \in \lset k \rset$ the vector $\p{d}_i \otimes \p{g}_i$ lies in 
    \begin{align*}
        \spanof 
        \left( \{
        {\p d}_i \otimes {\tilde{\p g}}_0, \dots, {\p d}_i \otimes {\tilde{\p g}}_{k-1}  
        \} \right). 
    \end{align*}
    By Lemma~\ref{lemma:decompositionofCsrm}, using the fact $\code{r-1}{m-1} \subset \code{r}{m-1}$ and the induction hypothesis, for each $i\in \lset k \rset$, the vector ${\p d}_i\otimes {\p g}_i \in \rowspaceof(G'_{r,m})$, while the vector ${\p d}_i\otimes {\tilde{\p g}}_i \in \rowspaceof(G_{r,m})$ for the same reasons. Thus, we have shown that $\rowspaceof(G_{r,m})\subseteq \rowspaceof(G'_{r,m})$. We can prove that  $\rowspaceof(G'_{r,m})\subseteq \rowspaceof(G_{r,m})$ using similar arguments. Thus,  $\rowspaceof(G_{r,m})= \rowspaceof(G'_{r,m})$. Recalling that $\rowspaceof(G_{r,m})=\code{r}{m}$ completes the proof. 
%
\end{IEEEproof}




\subsection{Reed-Muller and Dual Berman Codes as Special Cases}

We now show that the class of codes identified in~\cite{Ber_Cybernetics_II_67,BlN_IT_01,NaK_IT_23}, known as Dual Berman codes fall under the class of codes introduced in this work. 
It is known that the class of Dual Berman codes includes the family of RM codes.
Thus, this inclusion property carries over to our codes also.

We first recall the definition of the Dual Berman codes. For given integers $n\geq 2, m\geq 1$ and $r\in\{0,\hdots,m\}$, the Dual Berman codes $\dber{n}{ r,m}$ are subspaces of $\Fb_2^{n^m},$ defined recursively in the following manner in \cite{NaK_IT_23}.
\begin{align*}
    &\dber{n}{ m}{m}\triangleq \Fb_2^{n^m}, &&\\
    &\dber{n}{ 0}{m}\triangleq \{(c, \hdots, c)\in \Fb_2^{n^m}\colon c\in \Fb_2\},
\end{align*} 
and for $m\geq 2$ and $1\leq r\leq m-1,$ 
\begin{align}
    \dber{n}{r}{m}\triangleq 
    \{({ {\p d}_0}+{ {\p d}}_1|{ {\p d}}_0+{ {\p d}}_2|\hdots|{ {\p d}_0}+{ {\p d}}_{n-1}|{ {\p d}_0}) \colon \forall { {\p d}_0}\in \dber{n}{r}{m-1}
    \label{eqn:DualBermanRecDef}
    \forall {{\p d}}_l\in \dber{n}{r-1}{m-1}, {\forall l \in \{1,\hdots,n-1\}} \}.
\end{align}
It can be observed that $\RM(r,m)=\dber{2}{r}{m}$, following the Plotkin construction for $\RM$ codes. 

Now, let $\Cs=\Fb_2^n$ and $\{\p{e}_1,\dots,\p{e}_{n}\}$ be the standard basis for $\Fb_2^n$. 
Let the subcode $\Csub$ be generated as the rowspace of the matrix $G_{sub}$ with rows $\p e_j\in\Fb_2^n:j\in\{1,\hdots,n-1\}$
Note that $\Csub$ does not contain the vector $\allonelengthn$. We now show that $\code{r}{m}=\dber{n}{r}{m}$, up to a permutation of the coordinates.

Clearly, this is true for $r=0$ and $r=m$, as the definitions coincide in both of these cases to the repetition code of length $n^m$ and the $m$-fold product code $\Cs^{\otimes m
}=\Fb_2^{n^m}$, respectively. Thus, the claim is true for $m=1$. We argue the case for $m\geq 2$ (and for $r\in\{1,\hdots,m-1\}$) using induction on $m$, assuming that the claim is true for $m-1$. To see this, from Lemma~\ref{lemma:decompositionofCsrm}, observe that any codeword $\p c$ in $\code{r}{m}$ can be written as $\p c={\p d}_0\otimes\allonelengthn + \sum_{i=1}^{n-1}{\p d}_i\otimes {\p e}_i$ for some  ${\p d}_0\in\code{r}{m-1}$, and ${\p d}_i\in\code{r-1}{m-1}, \forall i\in \{1,\hdots,n-1\}$. 

Now, there exists (see \cite{Commutation_matrix_1979}) a permutation $\sigma$ of the coordinates $\{1,\hdots,n^m\}$, such that for any two vectors $\p a\in \Fb_2^{n^{m-1}}$ and $\p b\in\Fb_2^n$, the action of $\sigma$ on the coordinates of $\p a \otimes \p b$ results in $\p b\otimes \p a$. Abusing the notation slightly, we write, $\sigma(\p a \otimes \p b)=\p b\otimes \p a$. Thus, $\sigma(\p c)=\allonelengthn\otimes{\p d}_0 + \sum_{i=1}^{n-1}{\p e}_i\otimes {\p d}_i$. It is easy to see that $\sigma(\p c)=({ {\p d}_0}+{ {\p d}}_1|{ {\p d}}_0+{ {\p d}}_2|\hdots|{ {\p d}_0}+{ {\p d}}_{n-1}|{ {\p d}_0})$. We see that ${\p d}_0\in\code{r}{m-1}=\dber{n}{r}{m-1}$, and ${\p d}_i\in\code{r-1}{m-1}=\dber{n}{r-1}{m-1}, \forall i\geq 1$, where the equality in both expressions follows by the induction hypothesis. Thus, by definition of the code $\dber{n}{r}{m}$ as in (\ref{eqn:DualBermanRecDef}), it follows that $\sigma(\p c)\in\dber{n}{r}{m}$. Indeed, it is also easy to see by similar arguments that every codeword in $\dber{n}{r}{m}$ can be written as $\sigma(\p c)$, for $\p c= {\p d}_0\otimes\allonelengthn + \sum_{i=1}^{n-1}{\p d}_i\otimes {\p e}_i\in \code{r}{m}$. This completes the proof that $\dber{r}{r}{m}$ is identical to $\code{r}{m}$, where $\Cs=\Fb_2^{n}$, up to a permutation of the coordinates. 

\begin{remark}
Consider $\Cs=\Fb_2^2$, with the generator matrix of the subcode $G_{sub}=[0~1]$. Using these assignments, the codewords of $\code{r}{m}$ obtained in the form given by (\ref{eqn:codewordofCsrmdecomposition}) can be written as $\{\p d\otimes (1,1)+{\p d}'\otimes(0,1): \forall {\p d}\in\code{r}{m-1}, \forall {\p d}'\in\code{r-1}{m-1}\}$. It is easy to see that this is identical to the Plotkin construction of the code $\RM(r,m)$ (up to a permutation of the $2^m$ coordinates). Thus, from (\ref{eqn:codewordofCsrmdecomposition}), our construction of $\code{r}{m}$ can be seen as a generalization of the Plotkin construction of $\RM$ codes. 
\end{remark}


\subsection{Minimum Distance and Minimum Weight Codewords}

For a linear code $\cal C$, let $\Amin({\cal C})$ denote the set of all minimum weight codewords of $\cal C$, i.e., 
$$\Amin({\cal C})=\{\p c \in {\cal C}:\wt(\p c)=\dmin({\cal C})\}.$$ The following claim identifies the minimum distance of $\code{r}{m}$. 
\begin{lemma}
\label{lemma:mindist}
 The minimum distance of the code $\code{r}{m}$ is $\dmin(\code{r}{m})=\dmin(\Cs)^rn^{m-r}$. 
\end{lemma}
\begin{IEEEproof}
Let ${\p c}$ be a codeword in $\Amin(\code{r}{m})$. We express $\p c$ as the concatenation of $n^{m-1}$ $n$-tuples as $\p c=({\p c}_1|{\p c}_2|\hdots|{\p c}_{n^{m-1}})$.  By (\ref{eqn:codewordofCsrmdecomposition}) of Lemma~\ref{lemma:decompositionofCsrm}, we can then write  
\begin{align}
{\p c}=\sum_{i=0}^{k-1}{\p d}_i\otimes {\p g}_i,
\end{align}
for some unique codewords 
\begin{align*}
{\p d}_0&=(d_{0,1},\hdots,d_{0,n^{m-1}})\in\code{r}{m-1},  \text{ and} \\
{\p d}_i&=(d_{i,1},\hdots,d_{i,n^{m-1}})\in\code{r-1}{m-1}, \forall i\in \{1,\hdots,k-1\}.
\end{align*}
Observe that, for any position $t\in\{1,\hdots,n^{m-1}\}$, we have  
$${\p c}_t=\sum_{i=0}^{k-1}d_{i,t}\cdot{\p g}_i\in\Cs.$$ 
As the vectors $\{{\p g}_i:i\in\lset k \rset\}$ are precisely the (linearly independent) rows of $G$, we see that $\p 0 \neq {\p c}_t\in\Cs$ if at least one of the symbols $d_{i,t}:i\in\lset k \rset$ is non-zero (i.e., equal to $1\in\Fb_2$). 

Using this observation, and due to the fact that $\wt(\p g_0)=\wt(\allonelengthn)=n$, we immediately obtain the following bound on the weight of $\p c$. Let $S=\cup_{i=1}^{k-1}\supp({\p d}_i)$. Then,  
\begin{align}
    \label{eqn:lowerboundweightofcodewordofCsrm}
    \wt(\p c) = \sum_{t=1}^{n^{m-1}} \wt(\p{c}_t) \geq |\supp(\p d_0)\setminus S|\cdot n+|S|\cdot\dmin(\Cs). 
\end{align}
The rest of the proof is an induction argument based on~\eqref{eqn:lowerboundweightofcodewordofCsrm}, please see Appendix~\ref{app:proof:lemma:mindist} for details.
\end{IEEEproof}

The following claim shows the structure of the minimum weight codewords of $\code{r}{m}$, in a recursive manner. 

\begin{claim}
\label{claim:minwtcodewords}
The set of minimum weight codewords of $\code{r}{m}$, $\Amin(\code{r}{m})$, is the union of the below three sets:

\begin{enumerate}
    \item ${\cal A}_1=\{\tilde{\p d}\otimes \p g\colon \forall \tilde{\p d}\in \Amin(\code{r-1}{m-1}), \forall \p g\in \Amin(\Cs)\}$,
    \item ${\cal A}_2=\{{\p d}\otimes \allonelengthn+\tilde{\p d}\otimes \tilde{\p g}\colon \forall \tilde{\p d}\in \Amin(\code{r-1}{m-1}), \forall \p d\in \code{r}{m-1} ~\text{s.t}~\emptyset\neq \supp(\p d)\subsetneq  \supp(\tilde{\p d}),\forall \tilde{\p g}\in \Amin(\Cs)~\text{s.t.}~\allonelengthn+\tilde{\p g}\in \Amin(\Cs)\}$, and 
    \item ${\cal A}_3=\{{\p d}\otimes \allonelengthn\colon \forall{\p d}\in\Amin(\code{r}{m-1}) \}$. 
\end{enumerate}
\end{claim}
\begin{IEEEproof}
Please see Appendix~\ref{app:proof:claim:minwtcodewords}.
\end{IEEEproof}

The below corollary below follows by the observation that, for any $\tilde{\p g}\in \Amin(\Cs)$, we have $\wt(\allonelengthn+\tilde{\p g})=\wt(\tilde{\p g})$ if and only if $n=2\dmin(\Cs)$. 
\begin{corollary}
\label{corr:minwtcodewordsfornneq2dmin}
If $n\neq 2\dmin(\Cs)$, then $\Amin(\code{r}{m})={\cal A}_1\cup {\cal A}_3$ (where ${\cal A}_1$ and ${\cal A}_3$ are as in Claim \ref{claim:minwtcodewords}).   
\end{corollary}



We now provide an explicit identification of all the minimum weight codewords of $\code{r}{m}$, and prove that the minimum weight codewords of $\code{r}{m}$ span the code. We will use both these results for the {local graph search decoding}~\cite{Kam_TCOM_22} of the proposed codes in Section~\ref{sub:sec:LGS}.
 
\begin{lemma}
\label{lemma:tensorstructureofminwtcodewords}
Let $n\neq 2\dmin(\Cs)$. The minimum weight codewords of $\code{r}{m}$ can then be described as follows. 
\begin{align}
\nonumber
\Amin(\code{r}{m})=\bigcup_{J\subseteq \lset m \rset: |J|=r}\Amin(\code{r}{m},J),
\end{align}
where $\Amin(\code{r}{m},J)$ is defined as 
\begin{align*}
    \Amin&(\code{r}{m},J) 
    \triangleq \bigg\{\p h_0\otimes \p h_1\otimes\cdots\otimes\p h_{m-1}: 
    \p h_j\in \Amin(\Cs), \forall j\in J,~\text{and}~\p h_j=\allonelengthn, \forall j\notin J \bigg\}.
\end{align*}
\end{lemma}
\begin{IEEEproof}
We prove the lemma by induction on $m$. Clearly, for any $m$, $\code{0}{m}$ and $\code{m}{m}$ are the repetition code and the code $\Cs^{\otimes m}$, respectively. In the case of the repetition code, the minimum distance is $n^m$ and $\Amin(\code{0}{m})$ contains the codeword $\nosubscriptallone_{n^m}=\allonelengthn\otimes\cdots\otimes\allonelengthn$ ($m$ times). Thus, the lemma is true for $r=0$. For $\code{m}{m} = \Cs^{\otimes m}$, which is the product code, the minimum distance is precisely $\dmin(\Cs)^m$. 
We know that~\cite{Miller_78}
the codewords in $\Amin(\code{m}{m})$ precisely are 
$$\{\p h_0\otimes \p h_1 \otimes\cdots\otimes\p h_{m-1}: \forall \p h_j\in \Amin(\Cs), \forall j\}.$$ 
Thus, the lemma is true for $r=m$ also. These arguments also show that the lemma is true for $m=1$, for $r\in\{0,1\}$. 

Let $m\geq 2$. Assuming that the lemma is true for $m-1$ for each $r\in\lset m \rset$,  we show that the lemma is true for $m$, for $r\in\{1,\hdots,m-1\}$.

Firstly, we show that $\Amin(\code{r}{m},J)\subseteq \Amin(\code{r}{m})$.  Consider a vector $\p c=(\p h_0\otimes \p h_1\otimes\cdots\otimes\p h_{m-1})\in\Amin(\code{r}{m},J)$. We have that $\wt(\p c)=\prod_{j=0}^{m-1}\wt(\p h_j)=\dmin(\Cs)^rn^{m-r}=\dmin(\code{r}{m})$. We now show that $\p c\in\code{r}{m}$. 
If $\p h_{m-1}=\allonelengthn$, then $(\p h_0\otimes \p h_1\otimes\cdots\otimes\p h_{m-2})\in \Amin(\code{r}{m-1},J)$ by definition. Thus, by the induction hypothesis, $(\p h_0\otimes \p h_1\otimes\cdots\otimes\p h_{m-2})\in\code{r}{m-1}$. By Lemma \ref{lemma:decompositionofCsrm}, we see that $\p c=(\p h_0\otimes \p h_1\otimes\cdots\otimes\p h_{m-2})\otimes \allonelengthn \in \code{r}{m}$. 

On the other hand, if $\p h_{m-1}\in\Amin(\Cs)$, then $(\p h_0\otimes \p h_1\otimes\cdots\otimes\p h_{m-2})\in \Amin(\code{r-1}{m-1},J\setminus \{m-1\})$ by definition. Therefore, by the induction hypothesis, $(\p h_0\otimes \p h_1\otimes\cdots\otimes\p h_{m-2})\in \code{r-1}{m-1}$. 
Again, by invoking Lemma \ref{lemma:decompositionofCsrm}, we see that $\p c=(\p h_0\otimes \p h_1\otimes\cdots\otimes\p h_{m-2})\otimes \p h_{m-1} \in \code{r}{m}$. This completes the proof that $\Amin(\code{r}{m},J)\subseteq \Amin(\code{r}{m})$

We now show that any codeword in $\Amin(\code{r}{m})$ lies in $\Amin(\code{r}{m},J)$ for some $J \subset \lset m \rset$ such that $|J|=r$. By Corollary \ref{corr:minwtcodewordsfornneq2dmin}, the codewords in  $\Amin(\code{r}{m})$ are precisely comprised of the sets ${\cal A}_1$ and ${\cal A}_3$, as given in Claim \ref{claim:minwtcodewords}. 
Any codeword $\p c \in {\cal A}_1$ can be obtained as $\p c=\tilde{\p d}\otimes \p g$,  for some $\tilde{\p d}\in \Amin(\code{r-1}{m-1})$ and some $\p g\in \Amin(\Cs)$. 
By the induction hypothesis, there exists some $J'\subseteq \lset m-1 \rset$ such that $|J'|=r-1$, such that $\tilde{\p d} \in \Amin(\code{r-1}{m-1},J')$. That is, 
\begin{align*}
\tilde{\p d}=\p h_0\otimes \p h_1\otimes\cdots\otimes\p h_{m-2},
\end{align*}
where $\p h_j\in \Amin(\Cs),$ for all $j\in J'$, and $\p h_j=\allonelengthn$, for all $j\in \lset m-1 \rset \setminus J'$. It then follows that $\tilde{\p d}\otimes \p g \in \Amin(\code{r}{m},J'\cup\{m-1\})$. 

Now consider an arbitrary codeword  $\p c \in {\cal A}_3$. We can then write $\p c={\p d}\otimes \allonelengthn$,  for some ${\p d}\in \Amin(\code{r}{m-1})$. By the induction assumption ${\p d}\in \Amin(\code{r}{m-1},J')$ for some $J'\subseteq \lset m-1 \rset$ such that $|J'|=r$. Thus, it holds that $\p c\in \Amin(\code{r}{m},J')$. This completes the proof.
\end{IEEEproof}

The below corollary follows from Lemma \ref{lemma:tensorstructureofminwtcodewords}, by simple counting arguments. 
\begin{corollary} \label{corr:num_min_wt_codewords}
Let $n \neq 2\dmin(\Cs)$. The number of minimum weight codewords in $\code{r}{m}$ is $\binom{m}{r}|\Amin(\Cs)|^r$.
\end{corollary}

We now show that the minimum weight codewords of $\code{r}{m}$ span the code if the same is true for $\Cs$.  
\begin{claim} \label{claim:span_min_wt_codewords}
Let the code $\Cs$ be such that $n \neq 2\dmin(\Cs)$ and $\Cs=\spanof(\Amin(\Cs))$. Then, $\code{r}{m}=\spanof(\Amin(\code{r}{m}))$. 
\end{claim}
\begin{IEEEproof}
Firstly, for any $m$, the lemma holds for case of $r=0$ as $\code{0}{m}$ is just the repetition code. If $r=m$, then the code $\code{m}{m}=\Cs^{\otimes m}$ is generated by $G_1^{\otimes m}$, where $G_1$ is any generator matrix for $\Cs$. Since we know that $\Amin(\Cs)$ spans $\Cs$, the rows of $G_1$ can be chosen to be any $k$ linearly independent codewords from $\Amin(\Cs)$. This completes the proof for $r=m$. This also implies that the lemma holds for $r\in\{0,1\}$, when $m=1$.  
We now proceed by induction on $m$. For some $m\geq 2$, assume that the lemma is true for $m-1$, for all $r\in\lset m\rset$. We will show that it is true for $m$, for all $r\in\{1,\hdots,m-1\}$. 

From Lemma \ref{lemma:decompositionofCsrm}, we can write any codeword $\p c$ of $\code{r}{m}$ as $\p c=\sum_{i=0}^{k-1}{\p d}_i\otimes {\p g}_i$ for some  ${\p d}_0\in\code{r}{m-1}$, and ${\p d}_i\in\code{r-1}{m-1}, \forall i\in \{1,\hdots,k-1\}$. Recall that $\p g_i: i\in\{0,\hdots,k-1\}$ denote the rows of the generator matrix $G$ and $\p g_0=\allonelengthn$. 
Assume that $G_1=\begin{bmatrix}\p h_1^T &\p h_2^T &  \hdots & \p h_k^T \end{bmatrix}^T$ (where $\p h_i\in\Amin(\Cs), \forall i$ are linearly independent).

Now, $\p d_0\in\spanof(\Amin(\code{r}{m-1}))$ by the induction hypothesis. Thus, it holds that $\p d_0\otimes \p g_0=\p d_0\otimes\allonelengthn\in\spanof(\{\p h\otimes \allonelengthn:\forall \p h\in \Amin(\code{r}{m-1}\})$. It is easy to check that $\p h\otimes \allonelengthn\in \Amin(\code{r}{m})$ (following similar arguments as in proof of Lemma \ref{lemma:tensorstructureofminwtcodewords}). Thus, we have shown that $\p d_0\otimes \p g_0\in \spanof(\Amin(\code{r}{m}))$. 

Now, again by the induction hypothesis, $\p d_i\in\spanof(\Amin(\code{r-1}{m-1})), \forall i\geq 1$. 
Also observe that, for all $i\geq 1$, $\p g_i\in \Cs=\rowspaceof(G_1)$. 
Thus, $\p d_i\otimes \p g_i\in\spanof(\{\p h\otimes  \p h_j:\forall \p h\in \Amin(\code{r-1}{m-1}), \forall j\in\{1,\hdots,k\}\})$. 
It is easy to check that $\wt(\p h\otimes \p h_j)=\dmin(\code{r}{m})$. Further $\p h_j\in \rowspaceof(G)$. 
Thus, $\p h\otimes \p h_j\in\Amin(\code{r}{m})$. Thus, $\p d_i\otimes \p g_i\in \spanof(\Amin(\code{r}{m})), \forall i\geq 1$. 
Combining this with the fact that $\p d_0\otimes \p g_0\in \spanof(\Amin(\code{r}{m}))$, we have shown that $\p c=\sum_{i=0}^{k-1}{\p d}_i\otimes {\p g}_i\in \spanof(\Amin(\code{r}{m}))$. 
Along with the fact $\spanof(\Amin(\code{r}{m}))\subseteq \code{r}{m}$, the proof is complete.  
\end{IEEEproof}


\section{Fast ML Decoding of First-Order Codes} \label{sec:first_order_decoding}

We consider maximum-likelihood (ML) decoding of the first-order codes $\code{1}{m}$ in binary-input memoryless channels. 
The naive ML decoder for $\code{1}{m}$ has complexity order
\begin{equation*}
N \, 2^{\dim\left(\code{1}{m}\right)} = N \, 2^{1 + m(k-1)} = 2N^{1+\alpha},
\end{equation*}
where $N=n^m$ is the block length of the code and 
$$\alpha = \frac{(k-1)}{\log_2 n}.$$
In contrast, we use the recursive structure of the code to implement the ML decoder with complexity $\order(\max\{N,N^\alpha\})$ when $\alpha \neq 1$ and complexity $\order(N \log N)$ if $\alpha=1$.
We also propose an efficient soft-output max-log-MAP decoder with the same complexity order as the fast ML decoder. 
The ideas used here are similar to the fast ML decoders of first-order RM codes~\cite{BeS_IT_86,AsL_IT_04,ASY_IT_20} (which make use of the fast Hadamard transform).

We assume that the codewords are transmitted through a binary-input memoryless channel $W(y|x)$, where $x \in \{0,1\}$ and $y$ belongs to an output alphabet $\mathcal{Y}$. 
Suppose the channel output is $\p{y}=(y_1,\dots,y_N) \in \mathcal{Y}^N$, the corresponding channel log likelihood ratios (LLRs) are 
\begin{equation*}
\ell_i = \log_e \left( \frac{W(y_i|0)}{W(y_i|1)} \right), ~~i=1,\dots,N.
\end{equation*} 
Let $\p{\ell}=(\ell_1,\dots,\ell_N)$ denote the vector of LLRs.
We know that the ML decoder will output 
\begin{equation*}
\arg \max_{\p{c} \in \code{1}{m}} \sum_{i=1}^{N} (-1)^{c_i} \ell_i,
\end{equation*} 
where $\p{c}=(c_1,\dots,c_N) \in \code{1}{m}$ and $c_i \in \{0,1\}$. Let us denote the bipolar representation $(-1)^{c_i}$ of the binary digit $c_i$ as $c_i^{b}$, i.e., $c_i^b=+1$ if $c_i=0$ and $c_i^b=-1$ if $c_i=1$. Similarly, for a codeword $\p{c}$ define its bipolar representation
\begin{equation*}
\p{c}^b = \left( c_1^b,\dots,c_N^b  \right) \in \{+1,-1\}^N.
\end{equation*} 
Observe that the ML decoder chooses the codeword whose bipolar representation $\p{c}^b$ has the largest correlation $\langle \p{c}^b,\p{\ell} \rangle = \sum_{i}c_i^b \ell_i$ with the LLR vector $\p{\ell}$. 

\subsection{Efficient ML Decoding} \label{sub:sec:first_order_ML}

We will now use the recursive structure of $\code{1}{m}$ to identify an efficient recursive implementation of the ML decoder. Using Lemma~\ref{lemma:decompositionofCsrm} and the fact that $\code{0}{m-1}$ is the repetition code of length $n^{m-1}$, we observe that
\begin{align*}
\code{1}{m} = \Big\{ \p{d}\otimes \allonelengthn + \sum_{i=1}^{k-1} b_i \alloneB \otimes \p{g}_i~:  
\p{d} \in \code{1}{m-1}, b_1,\dots,b_{k-1} \in \Fb_2 \Big\},
\end{align*}
where $\alloneB$ denotes the all-ones binary vector of length $n^{m-1}$.
Since $\sum_{i=1}^{k-1} b_i \alloneB \otimes \p{g}_i = \alloneB \otimes \left( \sum_{i=1}^{k-1} b_i \p{g}_i \right)$, and since $\{ \p{g}_1,\dots,\p{g}_{k-1} \}$ is a basis for $\Csub$ we see that 
\begin{align}
\code{1}{m} \!=\! \Big\{ \p{d} \otimes \allonelengthn + \alloneB \otimes \p{a} : \p{d} \in \code{1}{m-1}, \p{a} \in \Csub \Big\}. \label{eq:firsr_order_code_rec}
\end{align} 

In order to work with the bipolar representation of the codewords let us use the notation $\p{u} \odot \p{v}$ to denote the Hadamard product (or the Schur product) of two vectors $\p{u}, \p{v}$, i.e., if $\p{u}$ and $\p{v}$ are vectors of same length then $\p{u} \odot \p{v}$ is the component-wise product of these two vectors.
From~\eqref{eq:firsr_order_code_rec}, we deduce that the bipolar representation of a codeword $\p{c} \in \code{1}{m}$ will be of the form
\begin{align*}
\p{c}^b = \left(\p{d}^b \otimes \allonelengthn \right) \odot \left(\alloneB \otimes \p{a}^b \right),~ \p{d} \in \code{1}{m-1}, \p{a} \in \Csub.
\end{align*} 
Using $\p{d}=(d_1,\dots,d_{n^{m-1}})$ and $\p{a}=(a_1,\dots,a_n)$ to denote the components of $\p{d}$ and $\p{a}$, we see that 
\begin{align} \label{eq:first_order_ML:1}
\p{c}^b = \p{d}^b \otimes \p{a}^b = \left(d_1^b\p{a}^b, d_2^b\p{a}^b, \dots,d_{n^{(m-1)}}^b\p{a}^b \right).
\end{align} 
Let us similarly split the $n^m$-length LLR vector $\p{\ell}$ into $n^{m-1}$ subvectors of length $n$ each, $\p{\ell}=\left(\p{\ell}_1,\dots,\p{\ell}_{n^{(m-1)}} \right)$, where $\p{\ell}_j = \left(\ell_{j,1},\dots,\ell_{j,n} \right)$ for $j=1,\dots,n^{(m-1)}$. 
Then 
\begin{align*}
\langle \p{c}^b, \p{\ell} \rangle = 
\sum_{j=1}^{n^{m-1}} \sum_{i=1}^{n} d_j^b a_i^b \ell_{j,i} = \sum_{j=1}^{n^{m-1}} d_j^b \, \left( \sum_{i=1}^{n} a_i^b \ell_{j,i} \right)
\end{align*} 
which is the correlation between the two $n^{(m-1)}$-length vectors $\p{d}^b$ and $\left( \sum_{i=1}^{n} a_i^b \ell_{1,i},\dots,\sum_{i=1}^{n} a_i^b \ell_{n^{m-1},i} \right)$. Let us denote this latter vector as $\p{{\ell}}(\p{a})$. Then,
\begin{equation} \label{eq:first_order_ML:2}
\langle \p{c}^b, \p{\ell} \rangle 
= \langle \p{d}^b \otimes \p{a}^b, \p{\ell} \rangle 
= \langle \p{d}^b, \p{{\ell}}(\p{a})\rangle.
\end{equation} 
While $\p{c} \in \code{1}{m}$, we see that $\p{d} \in \code{1}{m-1}$.
This relation allows us to implement the ML decoder recursively, by calling the decoder for $\code{1}{m-1}$ in order to decode $\code{1}{m}$. 
For each of the $2^{(k-1)}$ choices of $\p{a} \in \Csub$ we use the decoder for $\code{1}{m-1}$ to find the codeword  $\p{d}^b$ that maximizes $\langle \p{d}^b,\p{\ell}(\p{a}) \rangle$. Then, among all these $2^{(k-1)}$ choices we pick the one with the maximum correlation.
This decoder, denoted as $\mathcal{D}_{\Cs,m}$, is described in Algorithm~1.

\begin{center}
\begin{algorithm}
\caption{Efficient ML Decoder $\mathcal{D}_{\Cs,m}$ for $\code{1}{m}$}
{\bf Input}: Vector of channel LLRs $\p{\ell} \in \Rb^{n^m}$

{\bf Output}: $(\p{c}^b_{ML},M)$, where $\p{c}^b_{ML}$ is the bipolar representation of the ML codeword and $M=\langle \p{c}^b_{ML}, \p{\ell} \rangle$ is the corresponding metric

\begin{algorithmic}[1]

\If{$m=1$}

\State{Use brute-force ML decoder to find $\p{c}^b_{ML}$,  $M$}

\Return{$\left( \p{c}^b_{ML}, M \right)$}

\EndIf

\State \% Recursion for $m > 1$


\State{$\p{d}^b_{\sf best} \gets \p{0}$}

\State{$\p{a}_{\sf best} \gets \p{0}$} 

\State{$M_{\sf best} \gets -\infty$}

\For{each $\p{a} \in \Csub$}

\State{Compute the entries of the vector $\p{\ell}(\p{a})$}

\State{$\left(\p{d}^b(\p{a}), M(\p{a}) \right) \gets \mathcal{D}_{\Cs,m-1}\left( \p{\ell}(\p{a})\right)$}

\If{$M(\p{a}) > M_{\sf best}$}

\State{$M_{\sf best} \gets M(\p{a})$} 

\State{$\p{a}_{\sf best} \gets \p{a}$} 

\State{$\p{d}^b_{\sf best} \gets \p{d}^b(\p{a})$}

\EndIf

\EndFor


\State{$\p{c}^b_{ML} \gets \p{d}^b_{\sf best} \otimes \p{a}^b_{\sf best}$} 

\Return{$(\p{c}^b_{ML},M_{\sf best})$}


\end{algorithmic}
\end{algorithm}
\end{center}

\subsubsection*{Complexity Analysis}

We will use recursion to compute the complexity of the proposed decoder. Let $f(m)$ denote the complexity of the decoder $\mathcal{D}_{\Cs,m}$. For simplicity we will ignore the comparison operations (such as step~11) and assignment operations (such as steps~5--7 and step~17) in computing the complexity. 

For the case $m=1$ we use the brute-force ML decoder. This involves computing the correlation between $\p{\ell}$ and each of the $2^k$ codewords of $\Cs$. The resulting complexity is $n2^k$.
Note that it is possible to reduce this complexity to $n2^{(k-1)}$ by exploiting the fact that for each codeword (in bipolar representation) $\p{c}^b$ its negative $-\p{c}^b$ is also a codeword. This is a consequence of $\allonelengthn$ being a codeword of $\Cs$. Hence, it is enough to compute $\langle \p{c}^b, \p{\ell} \rangle$ for exactly half the codewords (for instance, the codewords in $\Csub$) since $\langle -\p{c}^b, \p{\ell}\rangle = -\langle \p{c}^b, \p{\ell} \rangle$. Hence, we have $f(1)=n2^{(k-1)}$.

For $m > 1$, we perform steps~9 and~10 of Algorithm~1 $2^{(k-1)}$ times. Step~9 has complexity $(n-1)n^{m-1} \leq n^m$, and step~10 has complexity $f(m-1)$. Thus, we have
\begin{equation*}
f(m) \leq 2^{(k-1)}\left( n^m + f(m-1) \right).
\end{equation*} 
Using the fact $f(1)=n2^{(k-1)}$ and using $2^{(k-1)}=n^\alpha$ (i.e., $\alpha=\frac{(k-1)}{\log_2 n}$), and unfolding the recursion, we obtain
\begin{align}
f(m) &\leq \sum_{i=1}^{m} n^{i\alpha} n^{(m+1-i)} = n^{m+1} \sum_{i=1}^{m} n^{i(\alpha-1)} \label{eq:fm:complexity}\\
&= n^{m+1} \left( \frac{n^{(m+1)(\alpha-1)} - n^{\alpha-1} }{n^{\alpha-1} - 1} \right) ~~\text{ if } \alpha \neq 1. \nonumber
\end{align}
We assume that $\Cs$ is fixed (i.e., $n$ and $\alpha$ are constants), and we are interested in the behaviour of $f(m)$ as a function of the blocklength $N=n^m$ as $m$ increases.
From the above expression, it is evident that $f(m) = \order(N \log N)$ if $\alpha=1$ (which is the case for RM codes). For $\alpha > 1$ we have $f(m) = \order(N^\alpha)$, and for $\alpha<1$ we have $f(m) = \order(N)$.  
 
We conclude that the decoder has complexity order $\max\{N,N^\alpha\}$ if $\alpha \neq 1$, and complexity order $N \log N$ if $\alpha=1$.


\subsection{Efficient Max-Log-MAP Decoding} \label{sub:sec:max_log_MAP}

We now describe a soft-output decoder for the first-order codes.
Later in this paper we will use this decoder along with the projection operation for decoding second-order codes $\code{2}{m}$ through belief propagation. 
The recursive ML decoder described in the previous sub-section does not provide soft-outputs (a posteriori LLRs). 
An efficient implementation of the optimal MAP (maximum a posteriori probability) decoder for first-order RM codes has been described in~\cite{AsL_IT_04}.
This decoder (which operates in the probability domain, not in the log domain) can be extended to $\code{1}{m}$. However, this decoder uses non-trivial functions, viz., $\log(\cdot)$ and $\exp(\cdot)$. This increases the implementation cost and also leads to numerical instability 
when working with small or large values of probability.

As an alternative, we consider the max-log-MAP decoder that can be implemented in the log domain. An approximation of the max-log-MAP decoder for first-order RM codes was proposed in~\cite{JFMH_ICC_22}. This approximation first computes the exact max-log-MAP outputs for the message bits and uses this to approximately compute the max-log-MAP output of the coded bits. 
Instead of such a two-step approximation, we provide a low-complexity recursive algorithm for exact max-log-MAP decoding for $\code{1}{m}$ that directly computes the soft-outputs for all coded bits.

Let us use the notation and decomposition set up in Section~\ref{sub:sec:first_order_ML}, especially~\eqref{eq:first_order_ML:1} and~\eqref{eq:first_order_ML:2}. 
Note that~\eqref{eq:first_order_ML:1} partitions $\p{c}^b$ into $n^{m-1}$ subvectors each of length $n$. The $i^\tth$ symbol in the $j^\tth$ subvector is $d_j^ba_i^b$, where $i=1,\dots,n$ and $j=1,\dots,n^{m-1}$. Thus, we have indexed the coded bits in $\p{c}$ using the tuple $(j,i)$.
The max-log-MAP decoder outputs 
\begin{equation*}
L_{j,i} = \log_e \left( \max_{\substack{\p{c} \in \code{1}{m}:\\ c_{j,i}^b=+1}} \Pb[\p{c} | \p{y}] \right) 
- \log_e \left( \max_{\substack{\p{c} \in \code{1}{m}:\\ c_{j,i}^b=-1}} \Pb[\p{c} | \p{y}] \right)
\end{equation*} 
for $i=1,\dots,n$, $j=1,\dots,n^{m-1}$. 
Here, $L_{j,i}$ is the max-log approximation of the true log APP (a posteriori probability) ratio given by
\begin{equation*}
\log_e \left( \sum_{\substack{\p{c} \in \code{1}{m}:\\ c_{j,i}^b=+1}} \Pb[\p{c} | \p{y}] \right) 
- \log_e \left( \sum_{\substack{\p{c} \in \code{1}{m}:\\ c_{j,i}^b=-1}} \Pb[\p{c} | \p{y}] \right).
\end{equation*}  
Following a procedure similar to~\cite[equation~(63)]{HOP_IT_96} it is rather straightforward to show that
\begin{equation*}
L_{j,i} = \frac{1}{2} \left(  \max_{\substack{\p{c} \in \code{1}{m}:\\ c^b_{j,i}=+1}} \langle \p{c}^b,\p{\ell}\rangle - \max_{\substack{\p{c} \in \code{1}{m}:\\ c^b_{j,i}=-1}} \langle \p{c}^b,\p{\ell}\rangle  \right).
\end{equation*} 
Let us denote the two terms in the above expression as 
\begin{equation*}
L_{j,i}^{(+1)} = \max_{\substack{\p{c} \in \code{1}{m}:\\ c^b_{j,i}=+1}} \langle \p{c}^b,\p{\ell}\rangle ~\text{ and }~
L_{j,i}^{(-1)} = \max_{\substack{\p{c} \in \code{1}{m}:\\ c^b_{j,i}=-1}} \langle \p{c}^b,\p{\ell}\rangle.
\end{equation*} 

\begin{center}
\begin{algorithm}
\caption{$\texttt{partialLLRs}_{\Cs,m}$ for the code $\code{1}{m}$}
{\bf Input}: Vector of channel LLRs $\p{\ell} \in \Rb^{n^m}$

{\bf Output}: Two vectors consisting of partial LLRs,\\ 
$\p{L}^{(+1)}=(L_{j,i}^{(+1)}: j=1,\dots,n^{m-1},i=1,\dots,n)$, and
$\p{L}^{(-1)}=(L_{j,i}^{(-1)}: j=1,\dots,n^{m-1},i=1,\dots,n)$
\begin{algorithmic}[1]

\If{$m = 1$}

\State{Compute the partial LLRs using the brute-force method.} 

\Return{$(\p{L}^{(+1)},\p{L}^{(-1)})$}

\EndIf

\State \% Recursion for $m > 1$

\For{each $\p{a} \in \Csub$}

\State{Compute $\p{\ell}(\p{a})$}

\State{$( \p{L}^{(+1)}(\p{a}), \p{L}^{(-1)}(\p{a}) ) \gets \texttt{partialLLRs}_{\Cs,m-1}\left( \p{\ell}(\p{a}) \right)$}

\EndFor

\For{each $j=1,\dots,n^{m-1}$ and $i=1,\dots,n$}

\State{$L_{j,i}^{(+1)} \gets \max_{\p{a} \in \Csub} L_{j}^{(a_i^b)}(\p{a})$}

\State{$L_{j,i}^{(-1)} \gets \max_{\p{a} \in \Csub} L_{j}^{(-a_i^b)}(\p{a})$}

\EndFor

\Return{$\left( \p{L}^{(+1)}, \p{L}^{(-1)} \right)$}


\end{algorithmic}
\end{algorithm}
\end{center}

\begin{center}
\begin{algorithm}
\caption{Max-Log-MAP Decoder $\texttt{maxlogMAP}_{\Cs,m}$ for the code $\code{1}{m}$}
{\bf Input}: Vector of channel LLRs $\p{\ell} \in \Rb^{n^m}$

{\bf Output}: A vector $\p{L}=(L_{j,i}: j=1,\dots,n^{m-1},i=1,\dots,n)$
\begin{algorithmic}[1]

\State{$\left( \p{L}^{(+1)}, \p{L}^{(-1)} \right) \gets \texttt{partialLLRs}_{\Cs,m}\left(\p{\ell}\right)$}

\For{each $j=1,\dots,n^{m-1}$ and $i=1,\dots,n$}

\State{$L_{j,i} \gets \frac{1}{2} \left( L_{j,i}^{(+1)} - L_{j,i}^{(-1)} \right)$}

\EndFor

\Return{$\p{L}$}

\end{algorithmic}
\end{algorithm}
\end{center}

Using~\eqref{eq:first_order_ML:1} and~\eqref{eq:first_order_ML:2} and the fact  $c^b_{j,i} = d^b_j\, a^b_i$, it is easy to observe that
\begin{align*}
L_{j,i}^{(+1)} &= \max_{\substack{\p{c} \in \code{1}{m}:\\ c^b_{j,i}=+1}} \langle \p{c}^b,\p{\ell}\rangle \\
&= \max_{\substack{\p{d} \in \code{1}{m-1},\,\, \p{a} \in \Csub: \\ d^b_j a^b_i = +1}} \langle \p{d}^b, \p{\ell}(\p{a}) \rangle \\
&= \max_{\p{a} \in \Csub} \left( \max_{\substack{\p{d} \in \code{1}{m-1}: \\ d_j^b = a_i^b}} \langle \p{d}^b, \p{\ell}(\p{a}) \rangle \right).
\end{align*}  
Observe that the inner maximization is a computation that would be used to perform max-log-MAP decoding of the code $\code{1}{m-1}$ where the channel LLRs are $\p{\ell}(\p{a})$. Using the notation 
\begin{align} 
L_j^{(+1)}(\p{a}) &=  \max_{\substack{\p{d} \in \code{1}{m-1}: \\ d_j^b = +1 }} \langle \p{d}^b, \p{\ell}(\p{a}) \rangle, \text{ and} \label{eq:maxlogmap:recpos} \\
L_j^{(-1)}(\p{a}) &=  \max_{\substack{\p{d} \in \code{1}{m-1}: \\ d_j^b = -1 }} \langle \p{d}^b, \p{\ell}(\p{a}) \rangle, \label{eq:maxlogmap:recneg}
\end{align}
we see that 
\begin{equation} \label{eq:maxlogmap:pos}
L_{j,i}^{(+1)} = \max_{\p{a} \in \Csub} L_{j}^{(a_i^b)}(\p{a})
\end{equation} 
Similarly, we can show that  
\begin{equation} \label{eq:maxlogmap:neg}
L_{j,i}^{(-1)} = \max_{\p{a} \in \Csub} L_{j}^{(-a_i^b)}(\p{a})
\end{equation}

In our recursive algorithm, we use a procedure for the code $\code{1}{m-1}$ to compute~\eqref{eq:maxlogmap:recpos} and~\eqref{eq:maxlogmap:recneg} for each $\p{a} \in \Csub$. We then use these values in~\eqref{eq:maxlogmap:pos} and~\eqref{eq:maxlogmap:neg} to compute the \emph{partial LLRs} $L_{j,i}^{(+1)}$ and $L_{j,i}^{(-1)}$. 
This procedure is described in Algorithm~2.
Finally, we compute the max-log-MAP decoder output as $\left( L_{j,i}^{(+1)} - L_{j,i}^{(-1)}\right)/2$, see Algorithm~3.


\subsubsection*{Complexity Analysis}

As in Section~\ref{sub:sec:first_order_ML}, we will assume that $\Cs$ is fixed. We will denote the complexity of $\texttt{partialLLRs}_{\Cs,m}$ as $g(m)$, and derive a recursive expression for it.
The brute force algorithm for $m=1$ can be implemented with complexity $g(1)=n2^k = 2n n^\alpha \leq 3nn^\alpha$.
When $m>1$, we see that steps~5--8 have complexity at the most $2^{(k-1)}(n^m + g(m-1)) = n^\alpha(n^m + g(m-1))$,
and steps 9--12 have complexity $n^m2^k = 2n^m n^\alpha$. Hence,
\begin{align*}
g(m) &\leq n^\alpha(n^m + g(m-1)) + 2n^m n^\alpha \\
&= n^\alpha(3n^m + g(m-1))
\end{align*} 
Using $g(1) \leq 3nn^\alpha$, it is easy to see that
\begin{align*}
g(m) \leq 3 \sum_{i=1}^{m} n^{i\alpha} n^{(m+1-i)} = 3n^{m+1}\sum_{i=1}^{m} n^{i(\alpha-1)}.
\end{align*} 
Comparing this with~\eqref{eq:fm:complexity}, we see that the complexity order of \texttt{partialLLRs} is same as that of Algorithm~1, which is $\order(\max\{N,N^\alpha\})$ if $\alpha \neq 1$ and is equal to $\order(N \log N)$ if $\alpha=1$. Further, since the complexity of steps~2--4 of $\texttt{maxlogMAP}$ is $\order(N)$, we conclude that the proposed max-log-MAP decoder has overall complexity order identical to that of the fast ML decoder given in Algorithm~1.

\section{Decoding Second-Order Codes} \label{sec:second_order_decoding}

Borrowing ideas from the literature on decoding RM codes~\cite{SiP_PPI_92,Sak_ITW_05,YeA_IT_20,LHP_ISIT_20} we propose a belief-propagation (BP) decoder that exploits a {projection} property of $\code{r}{m}$, viz., puncturing a codeword in two carefully chosen sets of coordinates and taking the sum of these subvectors yields a codeword in $\code{r-1}{m-1}$. Since we know how to decode $\code{1}{m-1}$ efficiently we can exploit this structure for decoding $\code{2}{m}$. 
Similar ideas have been used to decode certain cyclic codes in~\cite{ZhH_Globecom_22}. 
We then use a local graph search algorithm~\cite{Kam_TCOM_22} to improve the performance of the BP decoder.

\subsection{Projection Operation}

In order to describe the projection operation for $\code{r}{m}$ we index the coordinates of the codewords using elements from $\lset n \rset^m= \{0,1,\dots,n-1\}^m$ instead of the natural index set $\{1,2,\dots,n^m\}$.
We then introduce a notation for puncturing and show how this puncturing operation can be used to obtain projection for $\code{r}{m}$ similar to the projection technique used in decoding of RM codes~\cite{SiP_PPI_92,Sak_ITW_05,YeA_IT_20} and the derivative descendants defined for cyclic codes~\cite{ZhH_Globecom_22}.

\subsubsection{Indexing using $m$-tuples}

To create a new indexing we simply replace the index $i \in \{1,\dots,n^m\}$ with the coefficients of the base-$n$ expansion of $i-1$, i.e.,
\begin{equation*}
i-1 = i_0 n^{m-1} +i_1 n^{m-2} + i_2 n^{m-3} + \cdots + i_{m-1}n^0,
\end{equation*} 
where $i_0,\dots,i_{m-1} \in \lset n \rset$. 
Here the coefficients of the expansion are represented using the $m$-tuple $\p{i} = (i_0,\dots,i_{m-1})$. 
From~\eqref{eq:bj_basis_vector} we know that the $\p{j}^\tth$ basis vector for the code $\code{r}{m}$ is the Kronecker product 
\mbox{$\p{b}_{\p{j}} = \p{g}_{j_0} \otimes \cdots \otimes \p{g}_{j_{m-1}}$}, where \mbox{$\p{j}=(j_0,\dots,j_{m-1}) \in \lset k \rset^m$} and $\wt(\p{j}) \leq r$.
Representing the entries of $\p{g}_{j_{\ell}}$ as $(g_{j_{\ell},0},\dots,g_{j_{\ell},n-1})$ we observe that the $\p{i}^\tth$ entry of $\p{b}_{\p{j}}$ is 
\begin{equation*}
b_{\p{j},\p{i}} = g_{j_0,i_0} \times \cdots \times g_{j_{m-1},i_{m-1}}= \prod_{\ell \in \lset m \rset} g_{j_{\ell},i_{\ell}}.
\end{equation*} 
Any codeword in $\code{r}{m}$ can be uniquely expanded in terms of the basis $\{\p{b}_{\p{j}} : \p{j} \in \lset k \rset^m, \wt(\p{j}) \leq r\}$, i.e., if $\p{c} \in \code{r}{m}$ then there exist coefficients ${a}_{\p{j}} \in \Fb_2$ such that
\begin{equation*}
\p{c} = \sum_{\substack{\p{j} \in \lset k \rset^m: \wt(\p{j}) \leq r}} a_{\p{j}} \p{b}_{\p{j}}. 
\end{equation*} 
Thus for any $\p{c} \in \code{r}{m}$ the $\p{i}^\tth$ entry of $\p{c}$ is given by 
\begin{equation} \label{eq:transform_like_expansion}
c_{\p{i}} = \sum_{\substack{\p{j} \in \lset k \rset^m: \wt(\p{j}) \leq r}} a_{\p{j}} b_{\p{j},\p{i}} = \sum_{\substack{\p{j} \in \lset k \rset^m: \wt(\p{j}) \leq r}} a_{\p{j}} \prod_{\ell \in \lset m \rset} g_{j_{\ell},i_{\ell}}
\end{equation} 
for some choice of coefficients $a_{\p{j}} \in \Fb_2$.

\subsubsection{Puncturing} 

We intend to puncture a codeword \mbox{$\p{c} = (c_{\p{i}}:\p{i} \in \lset n \rset^m)$} onto the collection of indices $\p{i}$ where some of the coordinates of $\p{i}$ are frozen.
Let $\Fs \subset \lset m \rset$ denote the set of entries of $\p{i}$ that are going to be fixed, and let $\p{i}_{\Fs}=(i_{\ell}:\ell \in \Fs)$ be the corresponding subvector of $\p{i}$. If $f=|\Fs|$ then the length of $\p{i}_{\Fs}$ is $f$.
Let $\p{u} \in \lset n \rset^f$ be any $f$-length vector over $\lset n \rset$. Denote the collection of $\p{i} \in \lset n \rset^m$ such that the subvector $\p{i}_{\Fs}$ equals $\p{u}$ by $\Hs$, i.e.,
\begin{equation*}
\Hs = \left\{ \p{i} \in \lset n \rset^m~:~\p{i}_{\Fs} = \p{u} \right\}.
\end{equation*}  
For any codeword $\p{c}$ we will denote the subvector of $\p{c}$ obtained by puncturing to the indices $\Hs$ as 
\begin{equation*}
\Ps_{\Hs}(\p{c}) = \left( {c}_{\p{i}}~:~\p{i} \in \Hs \right).
\end{equation*}
The length of $\Ps_{\Hs}(\p{c})$ is $|\Hs| = n^{(m-f)}$.
This punctured vector can be indexed using $\lset n \rset^{(m-f)}$ as follows. 
For any $\p{i}'=(i_0',\dots,i_{m-f-1}') \in \lset n \rset^{(m-f)}$, the $\p{i}'^\tth$ entry of $\Ps_{\Hs}(\p{c})$ is equal to the $\p{i}^\tth$ entry of $\p{c}$ where $\p{i}$ is defined as follows 
$$\p{i}_{\Fs}=\p{u} \text{ and } \p{i}_{\lset m \rset \setminus \Fs} = \p{i}'.$$ 

\subsubsection{Projection Operation}

We now show that if we a puncture a codeword $\p{c} \in \code{r}{m}$ onto two carefully chosen sets of indices $\Hs$ and $\widetilde{\Hs}$, then the sum of the corresponding subvectors $\Ps_{\Hs}(\p{c}) + \Ps_{\widetilde{\Hs}}(\p{c})$ belongs to a code of order $(r-1)$.
Let $\Fs \subset \lset m \rset$ be of size $f$, and let $\p{u},\p{\tilde{u}} \in \lset n \rset^f$ be two distinct vectors. Define 
\begin{align*}
\Hs = \left\{ \p{i} \in \lset n \rset^m : \p{i}_{\Fs} = \p{u} \right\},~~ 
\widetilde{\Hs} = \left\{ \p{i} \in \lset n \rset^m : \p{i}_{\Fs} = \p{\tilde{u}} \right\}. 
\end{align*} 
Since $\p{u} \neq \p{\tilde{u}}$ we see that $| \Hs \cap \widetilde{\Hs} |=0$. 
For $\p{i}' \in \lset n \rset^{(m-f)}$, define $\p{i},\p{\tilde{i}} \in \lset n \rset^m$ as follows 
\begin{align*}
\p{i}_{\Fs}=\p{u},~ \p{i}_{\lset m \rset \setminus \Fs} = \p{i}', \text{ and }
\p{\tilde{i}}_{\Fs}=\p{\tilde{u}},~ \p{\tilde{i}}_{\lset m \rset \setminus \Fs} = \p{i}'.
\end{align*} 
Then the $\p{i}'\,^\tth$ entry of the projection $\Ps_{\Hs}(\p{c}) + \Ps_{\widetilde{\Hs}}(\p{c})$ is equal to the sum of the $\p{i}^\tth$ and $\p{\tilde{i}}^\tth$ entries of $\p{c}$.

\begin{lemma} \label{lemma:projection}
Let $\Hs$ and $\widetilde{\Hs}$ be as defined above, and $r-1 \leq m-f$. For every $\p{c} \in \code{r}{m}$, $$\Ps_{\Hs}(\p{c}) + \Ps_{\widetilde{\Hs}}(\p{c}) \in \code{r-1}{m-f}.$$
\end{lemma}
\begin{IEEEproof}
Please see Appendix~\ref{app:proof:lemma:projection}.
\end{IEEEproof}

\begin{remark}
{RM codes enjoy a richer set of projection operations which include the projections described in Lemma~\ref{lemma:projection} as a strict subset, see~\cite[Lemma~1]{YeA_IT_20}.}
\end{remark}

For any fixed $f \in \{1,\dots,m-r+1\}$ there are multiple choices of $\Fs$, $\p{u}$ and $\p{\tilde{u}}$ that can be used to perform projection. There are $\binom{m}{f}$ choices of $\Fs$, and for each such $\Fs$ there are $\binom{n^f}{2}$ choices for the pair $\{\p{u},\p{\tilde{u}}\}$, leading to $\binom{m}{f} \times \binom{n^f}{2}$ possible projections for a given value of $f$. Specifically, for $f=1$, we have $m\binom{n}{2}$ such projections, which is $\order(\log N)$.

\subsection{Belief Propagation Decoding of Second-Order Codes}

A BP decoder for RM codes was proposed in~\cite{LHP_ISIT_20} that exploits 
the fact that projecting an RM code of order $r$ yields an RM code of order $(r-1)$. 
This decoder uses a factor graph~\cite{KFL_IT_01} with hidden variable nodes (that represent the projected codeword bits) and generalized check nodes (to utilize the constraint that projected vectors belong to an RM code of order $(r-1)$). 
We use an identical approach to decode $\code{2}{m}$.

\subsubsection{Code Constraints used in BP Decoding}

We rely on two structural properties of $\code{2}{m}$.
First, using Lemma~\ref{lemma:projection} with \mbox{$r=2$} and \mbox{$f=1$}, we know that there exist $m\binom{n}{2}$ projections each of which yields a codeword from $\code{1}{m-1}$. This projected codeword can be efficiently decoded using the soft-in soft-out max-log-MAP decoder from Section~\ref{sub:sec:max_log_MAP}. 
Thus our factor graph contains $m\binom{n}{2}$ generalized check nodes, one corresponding to each of these projection operations, and where the max-log-MAP decoder of $\code{1}{m-1}$ is used during BP decoding. 
Recall that the complexity order of this generalized check node operation is $\max\{n^{(m-1)},n^{\alpha(m-1)}\}$ if $\alpha \neq 1$ and $(m-1)n^{(m-1)}$ if $\alpha=1$. Here $\alpha = \frac{(k-1)}{\log_2 n}$. Denoting the length of $\code{2}{m}$ by $N=n^m$, and assuming that $n$ and $\alpha$ are constant, we see that the complexity order of this generalized check node operation is $\max\{N,N^\alpha\}$ for $\alpha \neq 1$ and $N \log N$ for $\alpha=1$. 

Second, 
every codeword of $\code{r}{m}$ belongs to the product code $\Cs^{\otimes m}$. 
%
%
This property can be restated as follows.
Consider any $\Fs \subset \lset m \rset$ with $|\Fs|=m-1$ and any $\p{u} \in \lset n \rset^{(m-1)}$. Let $\Hs = \{\p{i}~:~\p{i}_{\Fs}=\p{u}\}$. Then, for every $\p{c} \in \code{r}{m}$, the punctured codeword $\Ps_{\Hs}(\p{c}) \in \Cs$.
Note that there are $\binom{m}{m-1} n^{(m-1)} = mn^{(m-1)}$ choices of $\Hs$.
In order to exploit this structure we use $mn^{(m-1)}$ generalized check nodes in our factor graph, one corresponding to each such choice of $\Hs$ (this is in addition to the generalized check nodes corresponding to projections); see~\cite[Fig.~2 and~3]{Tan_IT_81}. 
We perform brute-force max-log-MAP decoding at these nodes, and this has complexity $n2^k = 2n^{(1+\alpha)}$.
Finally, note that these generalized check nodes are useful (and need to be included in the factor graph) only if $\Cs$ is non-trivial, i.e., only if $k<n$.

\begin{figure}[!t]
\centering
\includegraphics[width=0.75\columnwidth]{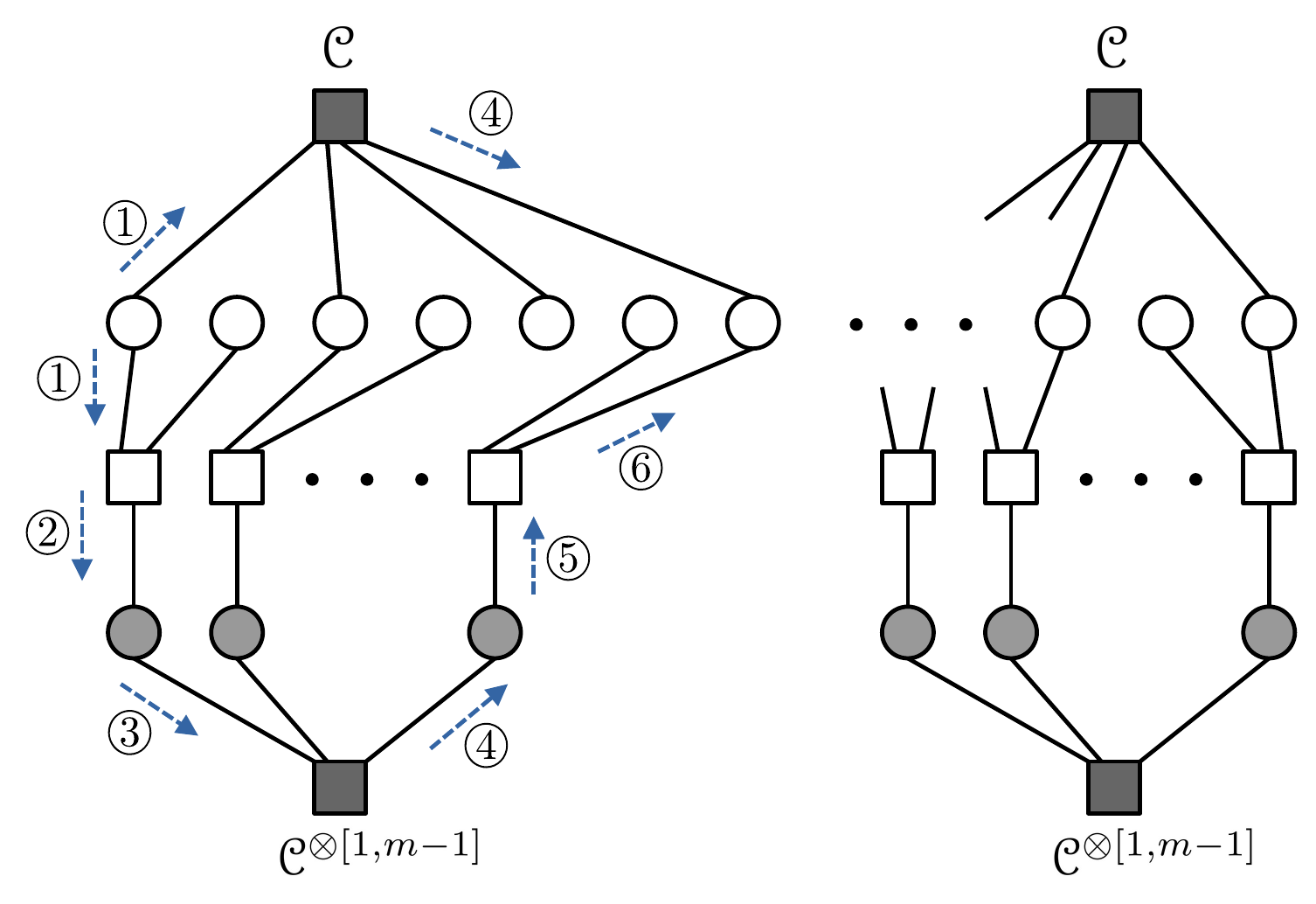}
\caption{Factor graph for BP decoding of $\code{2}{m}$. The numbers next to the dashed arrows represent the steps in each iteration. Empty circles are $\Vn$, empty squares are $\Chn$, filled circles are $\Vnh$, and filled squares are $\Chng$.}
\label{fig:BP_decoding}
\end{figure} 

\subsubsection{Factor Graph}

We now describe the nodes contained in our factor graph using a notation similar to~\cite{LHP_ISIT_20}, please see~Fig.~\ref{fig:BP_decoding}. There are four types of nodes in the graph:
\begin{enumerate}
\item[\emph{(i)}] $\Vn$: the collection of variable nodes corresponding to the code bits of $\code{2}{m}$. There are $n^m$ elements in $\Vn$ indexed using the set $\lset n \rset^m$.

\item[\emph{(ii)}] $\Chn$: these are the check nodes that capture the summation operation during projection. Each of these nodes is of degree $3$, two of its neighbors are in $\Vn$ and the third neighbor is a hidden variable node. Consider any $\Fs \subset \lset m \rset$ with $|\Fs|=1$ and distinct $u,\tilde{u} \in \lset n \rset$. For each pair $\p{i},\p{\tilde{i}} \in \Vn$ with 
\begin{equation*}
\p{i}_{\lset m \rset \setminus \Fs} = \p{\tilde{i}}_{\lset m \rset \setminus \Fs}, ~~\p{i}_{\Fs} = u, \text{ and } \p{\tilde{i}}_{\Fs} = \tilde{u},
\end{equation*} 
there exists a degree-$3$ check node in $\Chn$ with $\p{i},\p{\tilde{i}} \in \Vn$ being two of its neighbors. There are $m\binom{n}{2}n^{(m-1)}$ nodes in $\Chn$.

\item[\emph{(iii)}] $\Vnh$: this is the set of hidden variable nodes that represent all the code bits of all possible projected vectors $\Ps_{\Hs}(\p{c}) + \Ps_{\widetilde{\Hs}}(\p{c})$, i.e., these nodes represent the code bits of $\code{1}{m-1}$. 
The nodes in $\Vnh$ have degree $2$. One of their neighbors is a check node in $\Chn$ and the other neighbor is a generalized check node of type $\code{1}{m-1}$.
During BP decoding these nodes simply forward the incoming message on any of their edges to their other incident edge.

\item[\emph{(iv)}] $\Chng$: the set of generalized check nodes. This contains $m\binom{n}{2}$ nodes corresponding to the projected code $\code{1}{m-1}$, and $mn^{(m-1)}$ nodes corresponding to the code $\Cs$ (if $k < n$).  
Each node of type $\code{1}{m-1}$ is connected to exactly $n^{(m-1)}$ nodes in $\Vnh$. Every node of type $\Cs$ is connected to $n$ nodes in $\Vn$.  
We perform max-log-MAP decoding in all generalized check nodes to output the extrinsic bitwise LLR, i.e., the output of a generalized check node is equal to the max-log-MAP decoder's output minus the incoming message.
\end{enumerate} 

We are now ready to describe the BP decoder for $\code{2}{m}$.

\subsubsection{Belief-Propagation Decoding}

We use the standard BP algorithm on the factor graph with weighted extrinsic variable node update. To describe the algorithm we recall the notation used in~\cite{LHP_ISIT_20}. 
For any variable node $v$ (or a check node $c$) let $\nb v$ (respectively, $\nb c$) denote the neighbors of $v$ (respectively, $c$). Let $\p{\ell} \in \Rb^{n^m}$ denote the vector of channel LLRs, consisting of one scalar $\ell_v$ for each $v \in \Vn$. 
All the messages from check nodes (and generalized check nodes) to variable nodes are initialized to $0$, i.e., $\lambda_{c \to v}^{(0)}=0$, where the superscript denotes the iteration index. We set the iteration index $t=1$.

\step~1. 
The BP decoding iteration starts at the variable nodes $\Vn$. Each $v \in \Vn$ sends the following message to each check node and generalized check node $c \in \nb v$ in iteration $t$
\begin{equation} \label{eq:BP:Vn_update}
\lambda_{v \to c}^{(t)} = \ell_v 
+ 
\gamma \!\!\!\! \sum_{\substack{c' \in \nb v \cap \Chn: \\ c' \neq c}} 
\lambda_{c' \to v}^{(t-1)}
+
\gammag \!\!\!\! \sum_{\substack{c' \in \nb v \cap \Chng: \\ c' \neq c}} 
\lambda_{c' \to v}^{(t-1)},
\end{equation} 
where $\gamma, \gammag > 0$ are weights used to optimize the performance of the BP decoder. 
Note that all the generalized check nodes $c' \in \nb v \cap \Chng$ in the above sum correspond to the code $\Cs$.

\step~2.
The above operations are followed by updates in the check nodes $\Chn$ to generate messages to the hidden variable nodes $\Vnh$. For each $c \in \Chn$ and $v \in \nb c \cap \Vnh$, we compute 
\begin{equation} \label{eq:BP:Chn_update:1}
\lambda_{c \to v}^{(t)} = 2 \tanh^{-1} \left( \prod_{v' \in \nb c \cap \Vn} \tanh \left( \frac{1}{2} \lambda_{v' \to c}^{(t)} \right) \right).
\end{equation} 

\step~3.
The hidden variable nodes $v \in \Vnh$ then forward this incoming message to the generalized check nodes connected to them. 

\step~4.
At this point fresh LLRs are available as incoming messages along all the edges connected to every generalized check node. We perform max-log-MAP decoding at all \mbox{$c \in \Chng$} and compute the extrinsic LLRs by subtracting the incoming messages from the a posteriori LLRs. This gives us the messages $\lambda_{c \to v}^{(t)}$ for every $c \in \Chng$ and $v \in \nb c$.

\step~5.
The hidden variable nodes again forward the incoming messages from the generalized check nodes (of type $\code{1}{m-1}$) to their neighboring check nodes in $\Chn$. 

\step~6.
The check nodes $c \in \Chn$ then compute the messages to be forwarded to each of their neighboring variable nodes $v \in \nb c \cap \Vn$, 
\begin{equation} \label{eq:BP:Chn_update:2}
\lambda_{c \to v}^{(t)} = 2 \tanh^{-1} \left( \prod_{v' \in \nb c \setminus \{v\}} \tanh \left( \frac{1}{2} \lambda_{v' \to c}^{(t)} \right) \right).
\end{equation} 
We now increment the iteration index $t$ and continue with the next iteration. 

The BP algorithm is carried out till a specified maximum number of iterations, say $T_{\max}$, or till we converge to a valid codeword, whichever is earlier. The hard decisions on code bits $v \in \Vn$ are based on 
\begin{equation*} 
\ell_v 
+ 
\gamma \!\!\!\! \sum_{\substack{c' \in \nb v \cap \Chn}} 
\lambda_{c' \to v}^{(t)}
+
\gammag \!\!\!\! \sum_{\substack{c' \in \nb v \cap \Chng}} 
\lambda_{c' \to v}^{(t)}.
\end{equation*} 
We use a piecewise linear approximation~\cite[Table~II]{piecewise_linear_approx} to implement~\eqref{eq:BP:Chn_update:1} and~\eqref{eq:BP:Chn_update:2}.

\subsubsection{Complexity Analysis}

As before, we assume that $n$ and $\alpha$ are constants and analyse the complexity order as a function of length $N=n^m$ as $m$ varies.
The complexity of \step~1 is proportional to the total number of edges incident on all nodes in $\Vn$, which is equal to $mn^{(m+1)}=\order(N\log N)$.

The complexities of \step~2 and \step~6 are proportional to $|\Chn|$ since each of these nodes is of constant degree $3$. This is equal to $m\binom{n}{2}n^{(m-1)}=\order(N \log N)$.

In \step~4, there are $mn^{(m-1)}$ generalized check nodes of type $\Cs$ and $m\binom{n}{2}$ nodes of type $\code{1}{m-1}$. The overall complexity order of this step is 
\begin{equation} \label{eq:BP:overall_complexity_order}
\begin{cases}
N \log N & \text{ if } \alpha < 1, \\
N \log^2 N & \text{ if } \alpha = 1, \\
N^\alpha \log N & \text{ if } \alpha > 1.
\end{cases}
\end{equation}
Considering all the steps, we observe that each iteration of the BP decoder has complexity order equal to~\eqref{eq:BP:overall_complexity_order}. 

\begin{remark}
The complexities of the fast ML decoder for the first-order codes and the BP decoder for second-order codes increase with the value of $\alpha=\frac{(k-1)}{\log_2 n} = \frac{k-1}{n} \times \frac{n}{\log_2 n} $. 
The decoding complexity increases slowly with $N$ only when either the base code $\Cs$ has a small block length $n$, or a small rate $\frac{k}{n} = \order\left( \frac{\log_2 n}{n} \right)$.
Note that $\alpha=1$ for RM codes, while for Dual Berman codes $\alpha=(n-1)/\log_2 n$ is an increasing function of $n$.  
\end{remark}

\subsection{Improving the Performance via Local Graph Search} \label{sub:sec:LGS}

In order to improve upon the decoding error probability offered by the BP decoder we utilize the local graph search algorithm which was proposed in~\cite{Kam_TCOM_22} for RM codes.
This algorithm starts with the codeword output by the BP decoder and scans through a list of candidate codewords via a greedy technique, and chooses the best candidate from among this list.

\subsubsection{Search Algorithm}


\begin{figure}[!t]
    \centering
    \includegraphics[width=0.9\columnwidth]{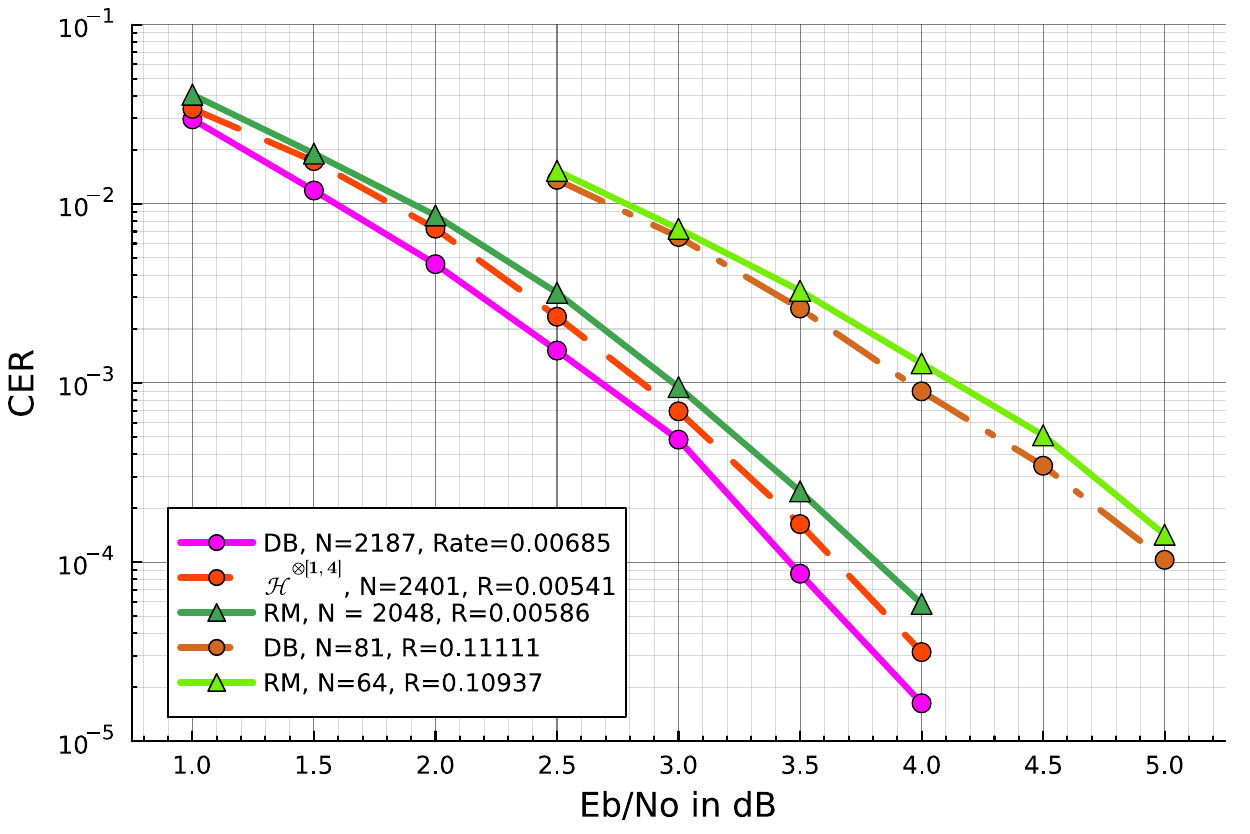}
    \caption{Performance of first-order codes under ML decoding. `DB' denotes Dual Berman codes, and $\mathcal{H}$ is the $[7,4,3]$ Hamming code.}
    \label{fig:first_order}
    \hfill 
\end{figure}
\begin{figure}[!t]
    \centering
    \includegraphics[width=0.9\columnwidth]{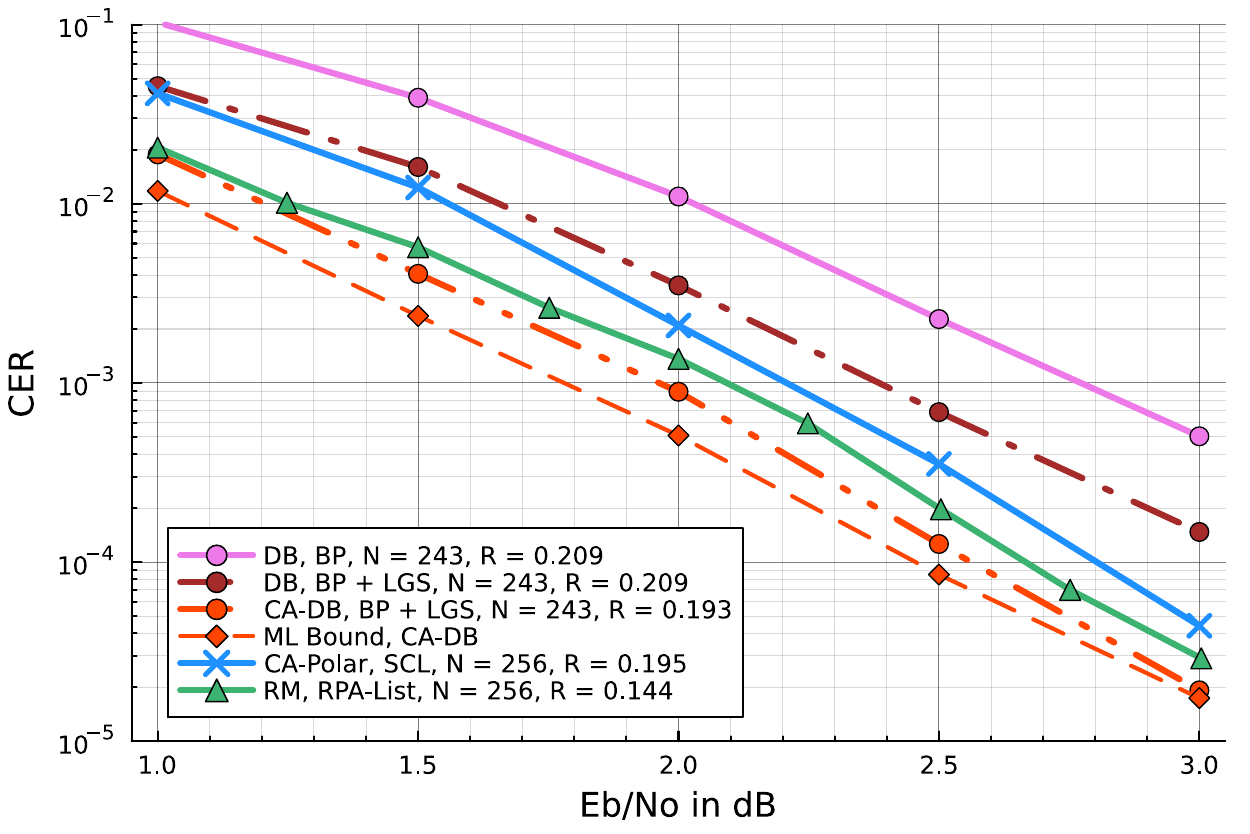}
    \caption{Performance of codes with block lengths close to $250$. Here, `DB' denotes the Dual Berman code $\dber{3}{2}{5}$, and `CA-DB' is its CRC-aided version.}
    \label{fig:length_250}
\end{figure}
\begin{figure}[!t]
    \centering
    \includegraphics[width=0.9\columnwidth]{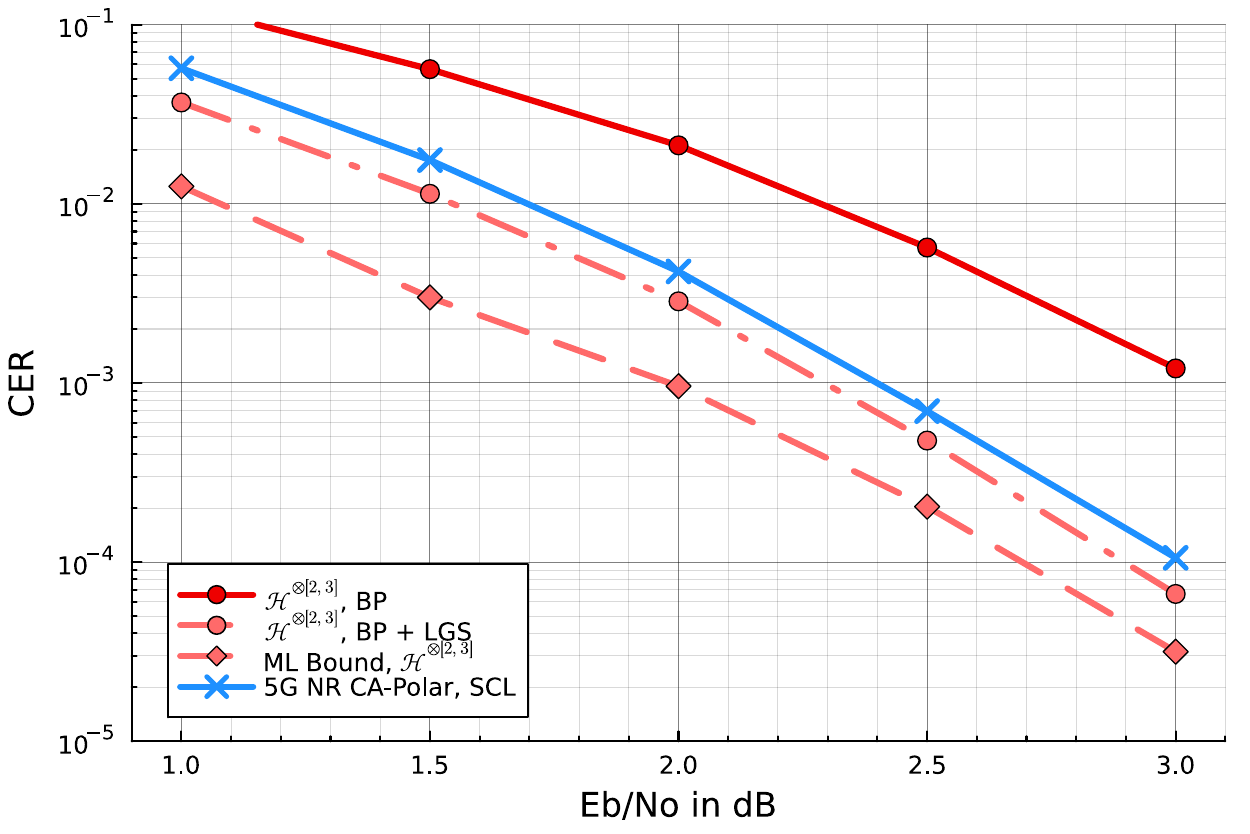}
    \caption{Comparison of $\mathcal{H}^{\otimes[2,3]}$ with 5G New Radio CRC-aided Polar code. Both codes are of length $343$ and dimension $37$.}
    \label{fig:length_343}
    \hfill 
\end{figure}
\begin{figure}[!t]
    \centering
    \includegraphics[width=0.9\columnwidth]{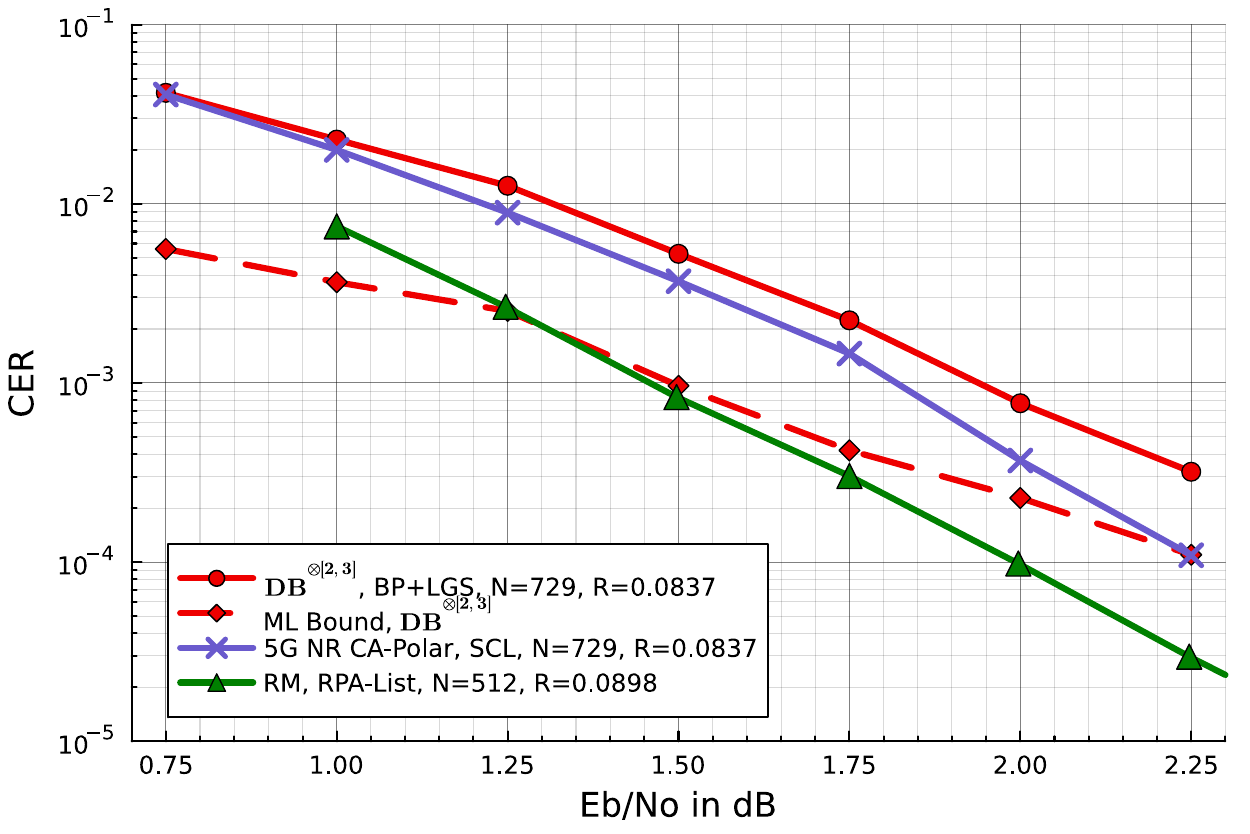}
    \caption{Codes with rates close to $0.085$. Here $\textrm{DB}$ denotes the $[9,5,3]$ Dual Berman code $\dber{3}{1}{2}$.}
    \label{fig:low_rate}
\end{figure}

The local graph search algorithm is best described through a graph $\mathscr{G}$ whose vertex set is the collection of all codewords in $\code{2}{m}$. 
An edge exists between two vertices if and only if the Hamming distance between the corresponding codewords is equal to the minimum distance of $\code{2}{m}$.
For any $\p{c} \in \code{2}{m}$ let $\nb \p{c}$ be the set of all neighbors of $\p{c}$ in $\mathscr{G}$.
Assume that $\p{c}_{\sf BP} \in \code{2}{m}$ is the codeword output by the BP decoder. Starting from $\p{c}_{\sf BP}$ we trace a path of some length, say $P_{\sf LGS}$, in $\mathscr{G}$ as follows. 
Initialize $p=0$ and set the $0^\tth$ node in the path as $\p{c}^{(0)} = \p{c}_{\sf BP}$. For $p=1,\dots,P_{\sf LGS}$, choose the $p^\tth$ node in the path as 
\begin{equation} \label{eq:next_step_LGS}
\p{c}^{(p)} = \arg \!\!\!\!\!\!\!\!\! \max_{\substack{\p{c} \in \nb \p{c}^{(p-1)}:\\ \p{c} \notin \{ \p{c}^{(0)},\p{c}^{(1)},\dots,\p{c}^{(p-1)} \}}} W\left( \p{y} | \p{c}  \right),
\end{equation}   
where $\p{y}$ is the channel output. 
That is, from among all the neighbors of $\p{c}^{(p-1)}$ that we have not visited so far we choose the codeword with the largest likelihood as $\p{c}^{(p)}$.
The algorithm terminates if $p=P_{\sf LGS}$ or if all the neighbors of $\p{c}^{(p-1)}$ have already been visited.
Once this path has been traversed, the algorithm outputs the most likely codeword from among all the codewords contained in the path.

We remark that $\mathscr{G}$ will be a connected graph if the minimum weight codewords of $\code{2}{m}$ span $\code{2}{m}$~\cite{Kam_TCOM_22}. 
In this case it is possible to reach any codeword in $\code{2}{m}$ from any other codeword via an appropriate path of sufficiently large length.
From Claim~\ref{claim:span_min_wt_codewords} we know that this is indeed the case if $\Cs=\Amin(\Cs)$, such as when $\Cs=\Fb_2^n$ or $\Cs$ is the $[7,4,3]$ Hamming code. 
Please note~\cite[Proposition~1]{Kam_TCOM_22} which shows that the search algorithm will return the ML codeword if instead of the greedy approach~\eqref{eq:next_step_LGS} we use an exhaustive breadth-first or depth-first search.
 
Note that it is possible for the output of the BP decoder to not be a codeword of $\code{2}{m}$. In general, we use the local graph search algorithm if the BP decoder output $\p{c}_{\sf BP}$ is indeed a codeword.
However, in the specific case of Dual Berman codes, i.e., when $\Cs=\Fb_2^n$, 
we can use a low complexity decoder $\mathcal{D}: \Fb_2^{n^m} \to \code{2}{m}$ described in~\cite[Algorithm~1]{NaK_IT_23} that has the following properties: \emph{(i)}~the output of the decoder lies in $\code{2}{m}$ for every input, and \emph{(ii)}~the decoder performs bounded distance decoding up to half the minimum distance of the code.
%
The complexity of $\mathcal{D}$ is $\order\left( N^{\left(1 + \log_2 n\right)} \right)$.
For Dual Berman codes we choose $\mathcal{D}(\p{c}_{\sf BP})$ as the starting node for the local graph search. 

\subsubsection{Complexity Analysis}

We will assume $n \neq 2\dmin(\Cs)$. 
Since $\code{2}{m}$ is a linear code, all the vertices in $\mathscr{G}$ have the same degree, which is equal to the number of minimum weight codewords in the code.
When \mbox{$n \neq 2\dmin(\Cs)$} we benefit from the small number of minimum weight codewords in $\code{2}{m}$, which is equal to $|\Amin(\Cs)|^2 \binom{m}{2} = \order\left( \log^2 N\right)$, see Corollary~\ref{corr:num_min_wt_codewords}. 

Computing the likelihood $W(\p{y}|\p{c})$ (or the log likelihood, for numerical stability) for a given neighbor has complexity $\order(N)$. 
We also need to check if a neighbor has already been visited. 
As in~\cite{Kam_TCOM_22}, we can use a red-black tree to store the collection of codewords visited so far. 
We can use the message bits of a systematically encoded codeword as the key to be stored in the red-black tree. The number of message bits for $\code{2}{m}$ is $\order(m^2) = \order(\log^2 N)$.
Verifying if a key of length $\order(\log^2 N)$ is present in the red-black tree has complexity $\order(\log^2 N \log P_{\sf LGS})$. 
Since these two steps have to be performed for every neighboring codeword and for up to $P_{\sf LGS}$ iterations, the overall complexity is 
$\order\left( P_{\sf LGS} \log^2 N \max\left\{N,\log^2 N \log P_{\sf LGS} \right\} \right)$.

\section{Simulation Results} \label{sec:simulations}

We now present the codeword error rate (CER) performance of a few recursive subproduct codes $\code{r}{m}$ with $r=1,2$ in the AWGN channel with BPSK modulation. 
We compare the performance of these codes with RM codes and CRC-aided Polar (CA-Polar) codes.
While recursive subproduct codes offer a wider range of block lengths and rates compared to RM codes, we have presented instances for which an RM code with a comparable length and/or rate exists.
{In all these presented results we observe that the CER of the recursive subproduct codes are either better or within $0.5$~dB (in terms of $E_b/N_o$) of the performances of RM and CA-Polar codes.} 

All first-order RM codes are decoded using the fast ML decoder. 
For other RM codes we consider the CER reported in~\cite{YeA_IT_20} under the list-decoding version of recursive projection aggregation (RPA). 
For CA-Polar codes we used the following two libraries: 
\emph{(i)}~the python library~\cite{RBV_GitHub} made available through~\cite{RBV_TVT_21}, and 
\emph{(ii)}~the Matlab library for 5G New Radio Polar codes for the uplink channel~\cite{Matlab_Polar}.

In our simulations of second-order codes $\code{2}{m}$ we have used $\Cs$ such that $\dmin(\Cs) < n/2$ and $\Amin(\Cs)$ spans $\Cs$. We know from Corollary~\ref{corr:num_min_wt_codewords} that the number of minimum weight codeword is $\order(m^2)=\order(\log^2 N)$, and from Claim~\ref{claim:span_min_wt_codewords} that these minimum weight codewords span $\code{2}{m}$, i.e., the graph $\mathscr{G}$ used in local graph search decoding is connected.

Fig.~\ref{fig:first_order} shows the CER of first-order codes under ML decoding. 
The two curves towards the right hand side of the figure are codes with short block lengths, viz., the Dual Berman code $\dber{3}{1}{4}=\code{1}{4}$ with $\Cs=\Fb_2^3$ (with length $81$, rate $0.111$, $\dmin=27$) and $\RM(1,6)$ (length $64$, rate $0.109$, $\dmin=32$).
The other three curves in Fig.~\ref{fig:first_order} correspond to codes with large block lengths, these are: 
\begin{itemize}
\item[\emph{(i)}] $\dber{3}{1}{7}=\code{1}{7}$ with $\Cs=\Fb_2^3$ (length $2187$, rate $0.0068$, $\dmin=729$), 
\item[\emph{(ii)}] $\code{1}{4}$ with $\Cs=\mathcal{H}$ the $[7,4,3]$ Hamming code (length $2401$, rate $0.0054$, $\dmin=1029$), and 
\item[\emph{(iii)}] $\RM(1,11)$ (length $2048$, rate $0.0058$, $\dmin=1024$).    
\end{itemize}
The superior performance of the subproduct codes over RM codes could be because of the small number of minimum weight codewords. The number of minimum weight codewords in $\RM(1,m)$ is $2^{m+1}-2= \order(N)$, whereas for $\code{1}{m}$ it is $m|\Amin(\Cs)| = \order(\log N)$.

Fig.~\ref{fig:length_250} compares the performance of the second-order Dual Berman code $\dber{3}{2}{5}=\code{2}{5}$, where $\Cs=\Fb_2^3$, with $\RM(2,8)$ and a CA-Polar code. 
Note that $\dber{3}{2}{5}$ is a $[243,51,27]$ code.
The CA-Polar code uses an $8$-bit CRC, is of length $256$ and has been decoded using successive-cancellation list (SCL) decoder with list size $32$. This CA-Polar code was simulated using~\cite{RBV_GitHub}.
The RM code is decoded using RPA-List decoder with list size $16$, which has near ML performance. Fig.~\ref{fig:length_250} contains four curves corresponding to $\dber{3}{2}{5}$ and its CRC-aided version:
\begin{enumerate}[\itshape(i)]
\item The right most curve in Fig.~\ref{fig:length_250} corresponds to BP decoding of $\dber{3}{2}{5}$ with up to $T_{\max}=5$ BP iterations and using weight $\gamma=0.12$ in~\eqref{eq:BP:Vn_update}. There are no generalized check nodes of type $\Cs$ in this factor graph since $\Cs=\Fb_2^3$ is a trivial code. Hence, in~\eqref{eq:BP:Vn_update}, we have $\nb v \cap \Chng = \emptyset$ for all $v \in \Vn$.

\item We obtain an improvement in performance by using the local graph search along with the BP decoder (denoted as `BP+LGS' in the figure). We use $P_{\sf LGS}=512$.

\item We strengthen the code by appending a $4$-bit CRC to $47$ message bits
using the CRC polynomial $x^4+x+1$. These $51$ bits are then encoded to a codeword in the Dual Berman code.
We use the same BP decoder and local graph search algorithm as with $\dber{3}{2}{5}$ except for a minor modification in the graph search stage: we choose the decoder output from only among the candidates that satisfy the CRC. We use $P_{\sf LGS}=2^{13}$.

\item A lower bound on the CER of the ML decoder for the CRC-aided Dual Berman code is also shown. This was computed using the same method as in~\cite{DuS_IT_06,YeA_IT_20}.
\end{enumerate}
We observe that the CRC-aided $\dber{3}{2}{5}$ code has a lower CER than $\RM(2,8)$ for the same $E_b/N_o$.

In Fig.~\ref{fig:length_343} we compare $\mathcal{H}^{\otimes [2,3]}$, where $\mathcal{H}$ is $[7,4,3]$ Hamming code, with the CA-Polar from 5G New Radio (5G NR)~\cite{BCL_Survey_21}. 
The code $\mathcal{H}^{\otimes [2,3]}$ has parameters $[343,37,63]$. 
The code from 5G NR has the same length and rate, uses a $11$-bit CRC, and is decoded via SCL with list size $32$.
Please note that the 5G NR standards let us generate Polar codes of length other than a power of $2$ via repetition, shortening and puncturing of a mother Polar code.
Fig.~\ref{fig:length_343} shows the performance of $\mathcal{H}^{\otimes [2,3]}$ under BP decoding ($\gamma=0.03$, $\gammag=0.25$, $T_{\max}=60$), as well as BP and local graph search decoding ($P_{\sf LGS}=512$). The ML lower bound on the CER of $\mathcal{H}^{\otimes [2,3]}$ is also shown.

Fig.~\ref{fig:low_rate} compares codes with rates close to $0.085$. 
The second-order RM code of length $512$ (dimension $46$, $\dmin=128$, under RPA decoding with list size $16$ which has near ML performance) and the 5G NR CA-Polar code of length $729$ and dimension $61$ (protected with a $11$-bit CRC, under SCL decoding with list size $32$) are used as benchmarks.
The recursive subproduct code presented in this figure is $\code{2}{3}$, where $\Cs=\dber{3}{1}{2}$ is the $[9,5,3]$ Dual Berman code. This code is of length $729$, dimension $61$, and minimum distance $81$. It has exactly $108$ minimum weight codewords. BP decoding (using $\gamma=0.0175$, $\gammag=0.3$, $T_{\max}=200$) with local graph search decoding ($P_{\sf LGS}=2^{13}$) was used. Its performance is about $0.5$~dB away from that of the RM code when the CER is $10^{-3}$.
A lower bound on the ML decoding performance of $\code{2}{3}$ is also shown.

\section{Discussion} \label{sec:discussion}


We identified the basic parameters of a family of subproduct codes $\code{r}{m}$ that include RM codes as a special case. 
We presented decoding algorithms for first- and second-order codes that use a fast recursion and a projection operation, respectively.

Further investigations into more efficient decoders are necessary.
The local graph search algorithm provides considerable performance improvement over the BP decoder, offering a CER that is close to ML decoding for second-order codes. However, this technique is costly since we needed a large number of steps $P_{\sf LGS}$ to obtain a close-to-ML performance, as much as $2^{13}$ in Fig.~\ref{fig:low_rate}. 
As for the BP decoder itself, it will be helpful to optimize the BP iterations to obtain a faster convergence and better CER.
Such improvements could be useful in designing recursive approaches to decode higher order codes (similar to RPA~\cite{YeA_IT_20}).

Our BP decoder for $\code{2}{m}$ uses projections corresponding to \mbox{$f=1$} from Lemma~\ref{lemma:projection}.
It is possible to extend the factor graph in Fig.~\ref{fig:BP_decoding} by including additional generalized check nodes corresponding to \mbox{$f \geq 2$}. However, our simulations showed no perceptible improvement in CER when generalized check nodes with \mbox{$f=2$} were included in BP decoding.

While including a $4$-bit CRC improved the performance of $\dber{3}{2}{5}$ (see Fig.~\ref{fig:length_250}), the inclusion of such a short CRC did not provide improvements (in CER with respect to $E_b/N_o$) for the other second-order codes presented in Section~\ref{sec:simulations}. 
It will be interesting to ascertain if a sufficiently long CRC can improve the performance of all second-order codes, and to design efficient decoders that can exploit the CRC in all stages of decoding (and not just during local graph search).

\appendix

\subsection{Proof of Lemma~\ref{lemma:decompositionofCsrm}} \label{app:proof:lemma:decompositionofCsrm}

The proof follows by observing the structure of the generator matrix $G_{r,m}$. 
For any $1\leq l\leq r-1\leq m-1$, we observe that the collection of rows of $B_{l,m}$ can be partitioned into the following collections:
\begin{itemize}
    \item the rows of the matrix $B_{l,m-1}\otimes\allonelengthn=B_{l,m-1}\otimes\p g_0$, and 
    \item the rows of the matrix $B_{l-1,m-1}\otimes G_{sub}$.
\end{itemize}
Also we observe that the rows of $B_{m,m}$ are precisely those of $B_{m-1,m-1}\otimes G_{sub}$, and that $B_{0,m}=B_{0,m-1}\otimes \allonelengthn=B_{0,m-1}\otimes\p g_0$. Thus, we have the following facts.
\begin{itemize}
    \item Fact 1: If $r\in\{1,\cdots,m-1\}$, then the rows of $G_{r,m}$ can be partitioned into two collections: the rows of $G_{r,m-1}\otimes \p g_0$, and the rows of $G_{r-1,m-1}\otimes G_{sub}$.
    \item Fact 2: $G_{0,m}=G_{0,m-1}\otimes \p g_0$.
    \item Fact 3: The rows of $G_{m,m}$ are precisely those of $G_{m-1,m-1}\otimes G$. 
\end{itemize}
Recall that $\code{r}{m}$ is defined as the rowspace of $G_{r,m}$. Now, for any matrices $A$ and $B=\begin{bmatrix}\p b_1^T&\p b_2^T&\hdots&\p b_s^T\end{bmatrix}$, it can be observed that $\rowspaceof(A\otimes B)=\spanof(\{\p a\otimes\p b_i: \forall \p a\in \rowspaceof(A), \forall i\in\{1,\hdots,s\}\}).$ Applying this observation to Fact 1 above, we see that 
\begin{align}
\nonumber
\code{r}{m}&=\spanof(\{\p d_0\otimes\p g_0: \forall \p d_0\in \code{r}{m-1}\})
+\spanof(\{\p d_i\otimes \p g_i: 
\forall\p d_i\in \code{r-1}{m-1},\forall i\in \{1, \hdots,k-1\}\})\\
&=\spanof(\{\p d_0\otimes\p g_0: \forall \p d_0\in \code{r}{m-1}\}) 
+\sum_{i=1}^{k-1}\spanof(\{\p d_i\otimes \p g_i: ,\forall\p d_i\in \code{r-1}{m-1}\}). \label{eqn:u3}
\end{align}
However,
\begin{align}
    \spanof(\{\p d_0\otimes\p g_0:& \forall \p d_0\in \code{r}{m-1}\})
    \label{eqn:u1}
    =\{\p d_0\otimes\p g_0: \forall \p d_0\in \code{r}{m-1}\}.
\end{align}
Similarly, we have 
\begin{align}
    \sum_{i=1}^{k-1}\spanof(\{\p d_i\otimes \p g_i: ,\forall\p d_i\in \code{r-1}{m-1}\}) 
    =\left\{\sum_{i=1}^{k-1} \p d_i\otimes\p g_i: \forall \p d_i\in \code{r-1}{m-1}\right\}. \label{eqn:u2}
\end{align}
Using (\ref{eqn:u1}) and (\ref{eqn:u2}) in (\ref{eqn:u3}), we see that the statement (\ref{eqn:codewordofCsrmdecomposition}) is true for $r\in\{1,\hdots,m-1\}$. The case of $r=0$ and $r=m$ follow using similar arguments on Facts 2 and 3. 


We now prove the uniqueness of the decomposition in (\ref{eqn:codewordofCsrmdecomposition}).
We do this by a counting argument. The number of choices of ${\p d}_0$ in (\ref{eqn:codewordofCsrmdecomposition}) is $2^{\dim(\code{r}{m-1})}$. The number of choices for each ${\p d}_i$ is $2^{\dim(\code{r-1}{m-1})}$ for $i \in \{1,\dots,k-1\}$. 
Thus, we observe that the cardinality of the set to the R.H.S. of (\ref{eqn:codewordofCsrmdecomposition}) is at most $2^{\dim(\code{r}{m-1})+(k-1)\dim(\code{r-1}{m-1})}$. But, 
\begin{align*}
    \dim(\code{r}{m-1})+(k-1)\dim(\code{r-1}{m-1})
    &=\sum_{l=0}^r\binom{m-1}{l}(k-1)^l+\sum_{l=0}^{r-1}\binom{m-1}{l}(k-1)^{l+1}\\
    &=1+\sum_{l=1}^{r}\left(\binom{m-1}{l}+\binom{m-1}{l-1}\right)(k-1)^l\\
    &=\sum_{l=0}^r\binom{m}{l}(k-1)^l \\
    &=\dim(\code{r}{m}).
\end{align*}
Thus, the number of vectors to the L.H.S and R.H.S of (\ref{eqn:codewordofCsrmdecomposition}) must be equal. Hence this decomposition must be unique.

\subsection{Induction Argument in the Proof of Lemma~\ref{lemma:mindist}} \label{app:proof:lemma:mindist}

We now show $\dmin(\code{r}{m})=\dmin(\Cs)^rn^{m-r}$ by induction on $m$. For $m=1$, it is easy to see by the definition of $\code{0}{1}$ and $\code{1}{1}$, that their minimum distances are $n$ and $\dmin(\Cs)$, respectively. Further, for $r=0$, we have that $\code{0}{m}=\spanof(\allonelengthn^{\otimes m})$ and thus $\dmin(\code{0}{m})=n^m$, for any $m\geq 1$. Thus, the statement is true for $m=1$, and for the case of $r=0$ and any $m$. For $r=m$, the claim is true as $\dmin(\code{m}{m})=\dmin(\Cs^{\otimes m})=\dmin(\Cs)^m$, where the last equality is known from the property of the $m$-fold product code of $\Cs$. 

Assume that the statement is true for $\code{r'}{m-1}$, for some $m\geq 2$ and for each value of $r'\in \{1,\dots,m-1\}$. We will prove that it is true for $\code{r}{m}$ for any $r\in \{1,\hdots,m\}$. Suppose, in the decomposition of $\p c$ as per (\ref{eqn:codewordofCsrmdecomposition}), we have that at least one of the ${\p d}_i:i\in\{1,\hdots,k-1\}$ is non-zero. This means, in (\ref{eqn:lowerboundweightofcodewordofCsrm}), we have that $S\neq \emptyset$. By the induction hypothesis, $|S|\geq 
\dmin(\Cs)^{r-1}n^{m-r}$. Thus, using (\ref{eqn:lowerboundweightofcodewordofCsrm}), we have that $\wt(\p c)\geq |S|\cdot \dmin(\Cs)\geq \dmin(\Cs)^{r}n^{m-r}$. If, on the other hand, we have ${\p d}_i=\p 0, \forall i$ (i.e., $S=\emptyset$), then we must have $\p d_0\neq \p 0$, as $\p c$ is non-zero. This means, using the induction hypothesis,
\begin{align*}
\wt(\p c)
= \wt \left( \p{d}_0 \otimes \p{g}_0 \right) 
=\wt(\p d_0)\cdot n 
\geq \dmin(\Cs)^rn^{m-1-r}\cdot n 
=\dmin(\Cs)^rn^{m-r}.    
\end{align*}
Thus we have that, for any non-zero $\p c\in \code{r}{m}$, $\wt(\p c)\geq \dmin(\Cs)^rn^{m-r}$.

We now show that there exists a codeword of weight $\dmin(\Cs)^{r}n^{m-r}$ in $\code{r}{m}$. We know that there exists a codeword $\tilde{\p g}\in \Amin(\Cs)$ such that $\tilde{\p g}=\sum_{i\in I}{\p g}_i$, for some $I\subseteq\lset k\rset$ (recall that ${\p g}_0=\allonelengthn$). Consider some $\tilde{\p d}\in\Amin(\code{r-1}{m-1})$. Fix ${\p d}_i=\tilde{\p d}, \forall i\in I$, and ${\p d}_i=\p 0, \forall i\in\lset k\rset \setminus I$. Using these assignments in (\ref{eqn:codewordofCsrmdecomposition}), and using the fact that $\code{r-1}{m-1}\subset\code{r}{m-1}$, we see that $\tilde{\p d}\otimes \tilde{\p g}=\tilde{\p d}\otimes\sum_{i\in I}{\p g}_i$ is a valid codeword in $\code{r}{m}$. Further, $\wt(\tilde{\p d}\otimes \tilde{\p g})=\wt(\tilde{\p d})\cdot\wt(\tilde{\p g})=\dmin(\Cs)^{r-1}n^{m-r}\cdot\dmin(\Cs)=\dmin(\Cs)^rn^{m-r}$. This proves that $\dmin(\code{r}{m})=\dmin(\Cs)^rn^{m-r}$.

\subsection{Proof of Claim~\ref{claim:minwtcodewords}} \label{app:proof:claim:minwtcodewords}

    Firstly, it is not difficult to check that the sets ${\cal A}_1, {\cal A}_2$ and ${\cal A}_3$ are subsets of $\code{r}{m}$, using (\ref{eqn:codewordofCsrmdecomposition}). Further, the weights of the codewords in ${\cal A}_1$ and ${\cal A}_3$ are easily seen to be equal to $\dmin(\Cs)^rn^{m-r}=\dmin(\code{r}{m})$. We now consider ${\cal A}_2$. Let ${\p c}={\p d}\otimes \allonelengthn+\tilde{\p d}\otimes \tilde{\p g}\in {\cal A}_2$. Let 
    \begin{align}
    \label{eqn:I1I2supports}
        I_1=\supp(\tilde{\p d})~\text{and}~I_2=\supp(\p d).
    \end{align} 
    Note that $|I_1|$ is equal to $\dmin(\code{r-1}{m-1})=\dmin(\Cs)^{r-1} n^{m-r}$. 
    Further, by the condition in ${\cal A}_2$, we have $I_2\subsetneq I_1\subseteq \{1,\hdots,n^{m-1}\}$. Express $\p c$ as the concatenation of $n^{m-1}$ $n$-tuples as $\p c=({\p c}_1|{\p c}_2|\hdots|{\p c}_{n^{m-1}})$. For any $t\in I_2$,  ${\p c}_t=\allonelengthn+\tilde{\p g}$. For any $t\in I_1\setminus I_2$, we have ${\p c}_t=\tilde{\p g}$. Finally, for any $t\in \{1,\hdots,n^{m-1}\}\setminus I_1$, we have ${\p c}_t=\p 0$. Thus, 
    $$\wt(\p c)=|I_2|\wt(\allonelengthn+\tilde{\p g})+|I_1\setminus I_2|\wt(\tilde{\p g}).$$ 
    From the properties of $\tilde{\p g}$ given in ${\cal A}_2$, we note that $$\wt(\tilde{\p g})=\dmin(\Amin(\Cs_{sub}))=\dmin(\Cs)=\wt(\allonelengthn+\tilde{\p g}).$$ 
    Thus, $\wt(\p c)=|I_1|\dmin(\Cs)=\dmin(\Cs)^{r} n^{m-r}$. Thus, the sets ${\cal A}_1, {\cal A}_2$ and ${\cal A}_3$ are all composed of minimum weight codewords of $\code{r}{m}$. We now show that any codeword in $\Amin(\code{r}{m})$ falls in ${\cal A}_1 \cup {\cal A}_2 \cup {\cal A}_3$. 
    
    Let $\p c$ be a minimum weight codeword in $\code{r}{m}$. From Lemma \ref{lemma:decompositionofCsrm}, we can write $\p c={\p d}\otimes\allonelengthn+\sum_{i=1}^{k-1}{\p d}_i\otimes {\p g}_i,$ for some unique codewords ${\p d}\in\code{r}{m-1}$ and ${\p d}_i\in\code{r-1}{m-1}, \forall i\in \{1,\hdots,k-1\}$. We consider two cases.

    \textit{Case 1: For some $i\geq 1,$ suppose ${\p d}_i\in\code{r-1}{m-1}$ is non-zero.} 
    
    In this case, using (\ref{eqn:lowerboundweightofcodewordofCsrm}), we have $$|S|\geq \dmin(\code{r-1}{m-1})=\dmin(\Cs)^{r-1}n^{m-r},$$ as $S=\cup_{i=1}^{k-1}\supp({\p d}_i)$. Using Lemma \ref{lemma:mindist}, we have $\wt(\p c)=\dmin(\Cs)^rn^{m-r}$, and thus we see from (\ref{eqn:lowerboundweightofcodewordofCsrm}), that $|S|=\dmin(\Cs)^{r-1}n^{m-r}=\dmin(\code{r-1}{m-1})$, and that $\supp(\p d)\setminus S=\emptyset$ (i.e., $\supp(\p d)\subseteq S$).

    Now, the fact that $|S|=\dmin(\code{r-1}{m-1})$ implies that all non-zero ${\p d}_i$s must have identical support of size $\dmin(\code{r-1}{m-1})$. 
    In other words, all non-zero ${\p d}_i$s must be identically equal to some minimum weight codeword, say $\tilde{\p d}$, in $\code{r-1}{m-1}$. Let $I\subseteq \{1,\hdots,k-1\}$ be such that ${\p d}_i= \tilde{\p d}, \forall i\in I$ and ${\p d}_i= {\p 0}, \forall i\in \{1,\hdots,k-1\}\setminus I$. 
    Thus, we can write 
    $$\p c ={\p d}\otimes\allonelengthn + \sum_{i=1}^{k-1}{\p d}_i\otimes {\p g}_i={\p d}\otimes\allonelengthn + \tilde{\p d}\otimes\left(\sum_{i\in I}{\p g}_i\right),$$ 
    where $\tilde{\p d}\in\Amin(\code{r-1}{m-1})$, and $\supp(\p d)\subseteq \supp(\tilde{\p d})$. Let $\tilde{\p g}=\sum_{i\in I}{\p g}_i\in\Cs_{sub}$. 
    Note that as $|I|\neq 0$, $\tilde{\p g}\neq \p 0$ (as ${\p g}_i$s are linearly independent). 
    
    Now, we consider three subcases.
    
    \underline{Subcase 1a: $\p d=\p 0$.} Then, $\dmin(\Cs)^rn^{m-r}=\wt(\p c)=\wt(\tilde{\p d})\wt(\tilde{\p g})=\dmin(\Cs)^{r-1}n^{m-r}\wt(\tilde{\p g})$, and thus we get that $\wt(\tilde{\p g})=\dmin(\Cs)$. Thus, if $\p d=\p 0$, we have that $\tilde{\p g}\in \Amin(\Cs)$, and hence $\p c=\tilde{\p d}\otimes\tilde{\p g}\in {\cal A}_1$. 
    
    \underline{Subcase 1b: $\p d=\tilde{\p d}$.} Then, we have $\p c=\tilde{\p d}\otimes(\allonelengthn+\tilde{\p g})$. Thus, in this case, we have $\dmin(\Cs)^rn^{m-r}=\wt(\p c)=\wt(\tilde{\p d})\wt(\allonelengthn+\tilde{\p g})=\dmin(\Cs)^{r-1}n^{m-r}\wt(\allonelengthn+\tilde{\p g})$, which means  $\wt(\allonelengthn+\tilde{\p g})=\dmin(\Cs)$. Thus, in this case too,  $\p c=\tilde{\p d}\otimes(\allonelengthn+\tilde{\p g})\in {\cal A}_1$. 

    \underline{Subcase 1c: Suppose $\supp(\p d)\subsetneq \supp(\tilde{\p d})$ and $\supp(\p d)\neq \emptyset.$} Using the definitions of $I_1$ and $I_2$ as in (\ref{eqn:I1I2supports}), observe that $|I_1|=\dmin(\code{r-1}{m-1})=\dmin(\Cs)^{r-1}n^{m-r}$. Further, $|I_1\setminus I_2|\neq 0$, and $|I_2|\neq 0$ by the assumption of this subcase. Then, using a similar analysis as before, we obtain 
    \begin{align*}
    \dmin(\Cs)^rn^{m-r}
    =\wt(\p c) =|I_2|\wt(\allonelengthn+\tilde{\p g})+(|I_1|-|I_2|)\wt(\tilde{\p g}).
    \end{align*} 
    Note that $\tilde{\p g}\in\Cs_{sub}$ and $\allonelengthn+\tilde{\p g}\in\Cs$, and both are non-zero, as each of them is a non-empty sum of linearly independent vectors. Thus, both $\tilde{\p g}$ and $\allonelengthn+\tilde{\p g}$ have weight at least $\dmin(\Cs)$. Therefore, it must be that, $\wt(\p c)\geq |I_1|\dmin(\Cs)=\dmin(\Cs)^rn^{m-r}$, with equality only when $\wt(\allonelengthn+\tilde{\p g})=\wt(\tilde{\p g})=\dmin(\Cs)$. This implies that, in this subcase, $\p c\in {\cal A}_2$. This concludes the subcases for Case 1. 

    \textit{Case 2: All ${\p d}_i:i\in\{1,\hdots,k-1\}$ are zero.} 
    
    In this case, $\dmin(\Cs)^rn^{m-r}=\wt(\p c)=\wt(\p d)\wt(\allonelengthn)=\wt(\p d)\cdot n.$ Thus $\wt(\p d)=\dmin(\Cs)^{r}n^{m-r-1}=\dmin(\code{r}{m-1})$. Thus, we see that $\p c={\p d}\otimes\allonelengthn\in {\cal A}_3$ in this case. This concludes the proof. 

\subsection{Proof of Lemma~\ref{lemma:projection}} \label{app:proof:lemma:projection}

For ease of exposition let us assume $\Fs=\{m-f,m-f+1,\dots,m-1\}$. The proof for other choices of $\Fs$ is similar. 

Since $\Fs=\{m-f,\dots,m-1\}$, the first $(m-f)$ positions of both $\p{i}$ and $\p{\tilde{i}}$ are equal to $\p{i}'$. The last $f$ positions of $\p{i}$ are equal to $\p{u}=(u_0,\dots,u_{f-1})$, while the last $f$ positions of $\p{\tilde{i}}$ are equal to $\p{\tilde{u}}=(\tilde{u}_0,\dots,\tilde{u}_{f-1})$.
Using these facts together with~\eqref{eq:transform_like_expansion} we see that the $\p{i}'\,^\tth$ entry of the projected vector $\Ps_{\Hs}(\p{c}) + \Ps_{\widetilde{\Hs}}(\p{c})$ is equal to 
\begin{align}
\sum_{\substack{\p{j} \in \lset k \rset^m: \\ \wt(\p{j}) \leq r}} 
a_{\p{j}} \!\!\! \prod_{\ell=0}^{m-f-1} \!\!\! g_{j_{\ell},i'_{\ell}}
\underbrace{
\left[ 
\prod_{\ell=m-f}^{m-1} \!\!\! g_{j_\ell,u_{\ell-m+f}} 
+ 
\prod_{\ell=m-f}^{m-1} \!\!\! g_{j_\ell,\tilde{u}_{\ell-m+f}} 
\right]
}_{\text{denote this as } \beta_{\p{j}}} 
= \sum_{\substack{\p{j} \in \lset k \rset^m: \\ \wt(\p{j}) \leq r}} 
a_{\p{j}} \beta_{\p{j}} \!\!\! \prod_{\ell=0}^{m-f-1} \!\!\! g_{j_{\ell},i'_{\ell}}. \label{eq:proof:projection:1}
\end{align} 
Note that $\beta_{\p{j}}$ is independent of $\p{i}'$. 

Suppose $\p{j}$ is such that $j_{\ell}=0$ for all $\ell \in \{m-f,\dots,m-1\}$, i.e., $\wt((j_0,\dots,j_{m-f-1})) = \wt(\p{j})$. Using the fact that $\p{g}_{0} = \allonelengthn$, we see that $\beta_{\p{j}}=0$. 
Hence,~\eqref{eq:proof:projection:1} contains contributions from only those $\p{j}$ that satisfy $\wt((j_0,\dots,j_{m-f-1})) < \wt(\p{j})$. Since the maximum weight of $\p{j}$ is $r$, the maximum weight of the subvector $(j_0,\dots,j_{m-f-1})$ is at the most $(r-1)$. 
Thus, the $\p{i}'\,^\tth$ component of the projected codeword $\Ps_{\Hs}(\p{c}) + \Ps_{\widetilde{\Hs}}(\p{c})$ is equal to
\begin{align*}
\sum_{\substack{(j_0,\dots,j_{m-f-1}) \in \lset k \rset^{m-f}:\\ \wt((j_0,\dots,j_{m-f-1})) \leq r-1 }} \!\!\! \prod_{\ell=0}^{m-f-1} \!\!\! g_{j_{\ell},i'_{\ell}}
\underbrace{
\left(
\sum_{\substack{(j_{m-f},\dots,j_{m-1}) \in \lset k \rset^f: \\ \wt(\p{j}) \leq r}} \!\!\!\!\!\!\! a_{\p{j}} \beta_{\p{j}}
\right)
}_{\text{denote this as } \alpha_{(j_0,\dots,j_{m-f-1})}} 
= \sum_{\substack{(j_0,\dots,j_{m-f-1}) \in \lset k \rset^{m-f}:\\ \wt((j_0,\dots,j_{m-f-1})) \leq r-1 }} \!\!\!\alpha_{(j_0,\dots,j_{m-f-1})} \prod_{\ell=0}^{m-f-1}  g_{j_{\ell},i'_{\ell}}. 
\end{align*} 
Comparing this last expression with~\eqref{eq:transform_like_expansion} we immediately conclude that $\Ps_{\Hs}(\p{c}) + \Ps_{\widetilde{\Hs}}(\p{c}) \in \code{r-1}{m-f}$.



\end{document}